\documentclass[12pt,tightenlines,floats,aps,prd,twoside,onecolumn,english,
	superscriptaddress,preprintnumbers,nofootinbib,showpacs,showkeys]{revtex4}

\usepackage{babel}
\usepackage{amsmath}
\usepackage{amsfonts}
\usepackage{amssymb}
\usepackage{dsfont}
\usepackage{graphicx}

\newcommand{\be}{\begin{equation}}
\newcommand{\ee}{\end{equation}}
\def\a{\alpha}
\def\b{\beta}
\def\g{\gamma}

\def\om{\omega}

\def\p{\partial}

\newcommand{\beq}{\begin{equation}}\newcommand{\eeq}{\end{equation}}\newcommand{\beqa}{\begin{eqnarray}}
\newcommand{\eeqa}{\end{eqnarray}}\newcommand{\w}{\wedge}
\newcommand{\nn}{\nonumber}

\newcommand{\cA}{\mathcal{A}}
\newcommand{\ou}[3]{\underset{#3}{\overset{#1}{#2}}}
\newcommand{\ua}{\uparrow}
\newcommand{\da}{\downarrow}

\begin{document}

\preprint{IGC-10/5-1}

\setcounter{page}{1}

\title{The Immirzi Parameter as an Instanton Angle}

\author{\firstname{Simone} \surname{Mercuri}}
\email{mercuri@gravity.psu.edu}
\affiliation{Institute for Gravitation and the Cosmos, The Pennsylvania State University,\\ 104 Davey Lab, Physics Department, University Park, PA 16802, USA}

\author{\firstname{Andrew} \surname{Randono}}
\email{arandono@perimeterinstitute.ca}
\affiliation{The Perimeter Institute for Theoretical Physics \\
31 Caroline Street North, Waterloo, ON N2L 2Y5, Canada}

\begin{abstract}
The Barbero-Immirzi parameter is a one parameter quantization ambiguity underpinning the loop approach to quantum gravity that bears tantalizing similarities to the theta parameter of gauge theories such as Yang-Mills and QCD. Despite the apparent semblance, the Barbero-Immirzi field has resisted a direct topological interpretation along the same lines as the theta-parameter. Here we offer such an interpretation. Our approach begins from the perspective of Einstein-Cartan gravity as the symmetry broken phase of a de Sitter gauge theory. From this angle, just as in ordinary gauge theories, a theta-term emerges from the requirement that the vacuum is stable against quantum mechanical tunneling. The Immirzi parameter is then identified as a combination of Newton's constant, the cosmological constant, and the theta-parameter. 
\end{abstract}

\pacs{04.50.Kd,04.60.Pp,04.20.Fy}

\keywords{de Sitter space, Barbero--Immirzi parameter, Instanton angle, Nieh--Yan density}

\maketitle

\section{Introduction}
The Barbero--Immirzi (BI) parameter is a real parameter that appears ubiquitously in the canonical formulation of quantum gravity known as Loop Quantum Gravity (LQG) \cite{AshLew04,Rovelli:Book,Thi07}. The BI parameter enters into the picture at the classical level, where it represents the coupling constant in front of a torsional term in the action, known as the Holst modification\cite{Holst}. Generally speaking, this modification of the action does not affect the classical equations of motion describing the dynamics of the gravitational field in the vacuum, but it does significantly alter the quantum theory, defining a one-parameter quantization ambiguity \cite{RovThi98}. Furthermore, many of the key features of LQG are dependent on the BI parameter. For example, the paramount achievement of LQG is the first principles derivation of a discrete spectrum for area and volume operators -- however, this spectrum is dependent on the value of the parameter. Similarly, the derivation of black-hole entropy in LQG from the counting of the quantum mechanical microstates yields an entropy that depends, as well, on the value of the BI parameter. Demanding that this entropy agrees with the semi-classical derivation of Bekenstein and Hawking the value of the BI parameter can be fixed \cite{Ashtekar:entropy} -- however, the derivation sheds dim light on why this particular value is special, and why the parameter is essential in the construction to begin with. Detractors of the theory may claim that the Planck scale discreteness predicted by LQG has more to do with the quantum dynamics of the Holst term, and less to do with generic features that one should expect from any theory of quantum general relativity. Indeed, one can study the Holst term alone as defining a quantum field theory -- although the theory is a topological field theory, at the kinematical level, the quantum features of Hilbert space appear to have many of the same features as the kinematical framework of full LQG \cite{Perez:HolstAction}.

Faced with these difficulties, it is of paramount importance to understand the true nature of the BI parameter. In fact, a better understanding of the parameter could shed light on why, or if, the parameter is needed in the theory, and what it has to do with Planck scale discreteness.

It has long been suspected that the BI parameter is analogous to the theta-parameter of Yang--Mills gauge theories \cite{GamObrPul99}. The similarities are difficult to ignore. However, there are some key differences that have allowed the BI parameter to elude such an interpretation. One of us, SM, has recently given an interpretation for the BI parameter as associated with the large sector of a local gauge group, in strict analogy with the $\theta$-ambiguity of Yang--Mills gauge theories \cite{Mer09p}. Specifically, further extending the Holst approach considering a true topological term in the action \cite{Mer09p,DatKauSen09},  the BI parameter becomes the coupling constant in front of such a parity violating term in the action, in analogy with the Pontryagin term of Yang--Mills gauge theories. In general, it does not affect the classical equations of motion, even in the presence of matter \cite{Mer09p} (see also \cite{Mer06}), but defines a one-parameter quantization ambiguity \cite{Mer08}, related to the large structure of the $SU(2)$ group \cite{Mer09p}. Here we offer an alternative interpretation based on a larger gauge group.

It has been known for some time that the two ingredients of Einstein--Cartan gravity, the spin connection and the tetrad, can be conveniently repackaged into a single connection valued in the de Sitter, anti-de Sitter, or Poincar\'{e} Lie algebra depending on the value of the cosmological constant\cite{MMoriginal}\cite{Stelle:1979va}\cite{Stelle:1979aj}\cite{Fukuyama:1984}. We will review this construction in section \ref{Section1}. This begs the question, {\it can gravity be interpreted as a gauge theory of this larger connection?} Without introducing new ingredients into the theory, the answer is negative --- it can easily be shown that the Einstein-Cartan equations of motion are not symmetric under the action of the larger gauge group, but only under the Lorentz subgroup. However, in light of the pervasive symmetry breaking mechanisms of the Standard Model, it is natural to think that gravity may be obtained as the symmetry broken phase of a more fundamental theory with exact local de Sitter, anti-de Sitter, or Poincar\'{e} gauge symmetry. Indeed, there are known mechanisms for doing this\cite{Stelle:1979va}\cite{Stelle:1979aj}, though these models are not without their deficiencies. Nevertheless, taking this principle seriously, one can ask if there are generic predictions that can be made based solely on the existence of such a theory, and not on the details therein. Recently one of us, AR, has shown that viewing gravity in this way can lead to new solutions to the Einstein--Cartan field equations, as there are some sectors of the symmetry reduced theory that retain a remnant of the full symmetry group, namely the vacuum solutions\cite{RandonodSSpaces}. This allowed for the construction of an infinite class of new solutions to the field equations, all of which are locally isomorphic to de Sitter space in regions where the metric is non-degenerate, but differing by topological properties. With respect to the symmetries of the more fundamental theory, all these solutions are related by a large de Sitter gauge transformation -- however, with respect to the reduced gauge symmetries of Einstein--Cartan gravity, the solutions are topologically distinct. In this paper, we again exploit the large sector of the de Sitter gauge group to argue for the existence of, in the de Sitter case, two instanton angles. The instantons in question, are quantum mechanical tunneling transitions between the degenerate vacua, which are classically identified with the aforementioned generalized class of de Sitter solutions. Upon symmetry breaking, one of the instanton angles can be related to the Immirzi parameter.

\section{Gravity as an (A)dS gauge theory}\label{Section1}
There are two main pieces of evidence, one observational and one theoretical, that the current formulation of gravity may be the remnant of a more fundamental theory based on the de Sitter group. The first is the observational discovery that the cosmological constant is, indeed, non-zero and positive. As a consequence, while the Universe expands and the ordinary matter content dilutes, it inevitably reaches a vacuum dominated phase that, at least locally, becomes more and more like de Sitter space as time progresses. We are currently in such a vacuum dominated era. The second is a theoretical construction of a de Sitter invariant theory of gravity known as the Macdowell-Mansouri mechanism, and its extensions, which combines the tetrad and the spin connection into a single connection that takes values in the de Sitter Lie algebra (we refer the reader to Appendix \ref{Appendix} for details). This can be easily seen by using a Clifford algebra notation, where the spin connection 1-form $\omega\equiv \frac{1}{4} \gamma^{[I}\gamma^{I]}\,\omega_{IJ}$ takes values in the bivector elements of the algebra, the tetrad, instead, takes values in the vector elements $e=\frac{1}{2}\gamma_I \,e^I$. Since the basis $\{\frac{1}{2}\g^{[I}\g^{J]}\,,\, \frac{1}{2}\g_5\g^K\}$ is a basis for the de Sitter Lie algebra, one can now combine the two fields into a single, de Sitter connection denoted $\mathcal{A}=\omega+\frac{1}{\ell}\gamma_5 e$, where $\ell$ is a parameter with the dimension of a length. Its curvature is $F_{\cA}=d\cA+\cA\wedge\cA$. The MacDowell--Mansouri action can then be written (explicit wedge products are dropped and the trace over the Clifford algebra is assumed)
\beq
S_{MM}=\frac{1}{\kappa} \int_M \star\,F_{\cA}\,F_{\cA} \label{MM}
\eeq
where $\kappa$ is a \emph{dimensionless} coupling constant and $\star=-i\g_5$. Expanding the curvature into its vector and bivector components, $F=R_{\omega}-\frac{1}{\ell^2}e\,e+\frac{1}{\ell}\gamma_5 T$, where $T=D_\omega e$ is the torsion, the action now reduces to (taking note that the trace over any odd number of Clifford matrices is zero)\footnote{Note that in this expression, the operator $\star$ acts like a dual on the internal indices. This implies, for example, that $Tr(\star(e\,e)\,R)=Tr(e\,e(\star R))=Tr(\star e\,e\,R)$. We refer the reader to the Appendix for more details.} :
\beq\label{eq1}
S_{MM}=\frac{1}{\kappa} \int_M \star \,R\,R -\frac{2}{\ell^2} \star e\,e \,R +\frac{1}{\ell^4}\star e\,e\,e\,e \,.
\eeq
Identifying the coupling constant with $\kappa=-\frac{16\pi G}{\ell^2}$ and the parameter $\ell
$ with $\sqrt{\frac{3}{\Lambda}}$ where $\Lambda$ is the cosmological constant, we recognize the resultant action as ordinary Einstein--Cartan gravity with a cosmological constant and a topological term proportional to the Euler characteristic.

Despite the formal analogy with Yang--Mills gauge theories, the action (\ref{eq1}) is fundamentally different. First note that the ``dual", denoted as $\star$, is the dual in the internal space to distinguish it from the operator $*$, which is the dual on the spacetime indices, appearing in the Yang--Mills theories. Furthermore, whereas the kinetic term of Yang--Mills retains the full gauge symmetry of the group it is based on, the operator $\star$ explicitly breaks the de Sitter invariance of the theory. As a consequence, despite the appearance of a de Sitter symmetric theory, the symmetry is broken at the level of the action, and the MacDowell--Mansouri mechanism could be (and often is) written off as simply a convenient way of repackaging variables, with no further significance.

On the other hand, there are known mechanisms for constructing an action along the lines of the MacDowell--Mansouri action that preserves the full local de Sitter symmetry and a natural, symmetry broken phase\cite{Stelle:1979va}\cite{Stelle:1979aj}\cite{Fukuyama:1984}\cite{Chamseddine:2010rv}. As is typical with symmetry breaking mechanisms, this requires the introduction of new dynamical fields, which dynamically relax to a vacuum state effectively breaking the symmetry. However, to our knowledge it is not known if these mechanisms are truly dynamical, in the sense that the symmetry broken phase is an energetically, or otherwise, \emph{preferred} ground state, that the system inevitably tends toward. Nevertheless, the existence of such quasi-dynamical symmetry breaking mechanisms, driving the initial $Spin(4,1)$ gauge theory to the $Spin(3,1)$ Einstein--Cartan gravity, can be seen as a strong motivation for taking seriously the $Spin(4,1)$ scenario. In this paper we will take the possibility of a more fundamental de Sitter symmetry seriously, and explore some of the consequences.

\subsection{A toy model}
Before discussing the emergence of the Barbero--Immirzi parameter, we present a model to illustrate some of the main features of the symmetry breaking mechanism, in particular the emergence of a theta-term. The model is discussed in full detail in \cite{Randono:Condensate}. As with the Stelle-West model, this model is quasi-dynamic in the sense that the order parameter does not, as of yet, have an associated kinetic term and realistic potential to spontaneously break the symmetry. However, the model is natural in the sense that employs familiar elements from condensed matter systems and the Standard Model of particle physics to break the symmetry through a physically motivated mechanism involving a highly symmetric fermion condensate. For our purposes it will illustrate some of the generic features one should expect from a physically realistic spontaneous symmetry breaking mechanism.

We begin with a multiplet of spinor fields $\psi_i$ where the index $i$ ranges from one to four and simply labels the spinors in the multiplet. The individual spinors in the multiplet are taken to be Dirac spinors, a term which we take in the most primitive sense to mean an element of a 4-component complex vector space subject to the inner product $\langle \phi |\psi \rangle_D =\bar{\phi} \psi=\phi^\dagger \gamma^0 \phi $. As such they furnish a representation of the universal cover of the conformal group in four dimensions (via the action of the Clifford algebra) with the de Sitter and anti-de Sitter groups as natural subgroups. 

The action we will consider takes the form
\beqa
S &=& \sigma \int_M DD\bar{\psi}_i\, DD\psi_i +S_{matter} \nn\\
&=& -\sigma \int_M \bar{\psi}_i F \,F \psi_i +S_{matter}\,.
\eeqa
We will leave the matter action undetermined, but we require that the full action is exactly symmetric under de Sitter gauge transformations.

It is useful to rewrite the action as follows
\beq
S=\sigma\int_{M} Tr(\psi_i \bar{\psi}_i\,F\,F) +S_{\rm matter}
\eeq
recognizing that $\psi_i \bar{\psi}_i$ is a complex $4\times 4$ matrix. The Fierz identity allow us to decompose the matrix into a Clifford basis as follows (here $\Gamma^{\hat{A}}$ is a normalized basis for the sixteen dimensional Clifford algebra):
\beqa
\psi_i\bar{\psi}_i&=&\sum_{\hat{A}} \bar{\psi}_i \Gamma_{\hat{A}}\psi_i \,\Gamma^{\hat{A}}\\
&=& -\frac{1}{4}\left(\bar{\psi}_i{\psi}_i\,1 -\bar{\psi}_i\star\psi_i\,\star-i\bar{\psi}_i\gamma_I{\psi}_i\,i\gamma^I+\bar{\psi}_i\star \gamma_I {\psi}_i\,\star\gamma^I+\frac{1}{2}\,i\bar{\psi}_i\gamma_{[I}\gamma_{J]}{\psi}_i\,i\gamma^{[I}\gamma^{J]}\right)\,. \nn\label{BilinearExpansion}
\eeqa

Let us now assume that there exists a preferred vacuum state where the auxiliary field is {\it isotropic} and {\it homogeneous} in both space and time. Isotropic means that the field does not isolate a preferred direction. Thus, on this solution, the vacuum expectation values of the vector, axial vector, and bivector elements vanish:
\beq
\langle\bar{\psi}_i \gamma^I \psi_i \rangle=0\quad \langle\bar{\psi_i}\star\gamma^I \psi_i \rangle=0 \quad \langle\bar{\psi}_i\gamma^{[I}\gamma^{J]} \psi_i \rangle=0\,.
\eeq
While homogeneous means that the remaining two expectation values are constant:
\beq
\langle\bar{\psi}_i\psi_i \rangle=-c_1=constant \quad \langle\bar{\psi}_i\star \psi_i \rangle=c_2=constant\,.
\eeq
As shown in \cite{Randono:Condensate} these constraints can be imposed in a de Sitter symmetric way with the specific form above emerging upon an appropriate choice of the local $Spin(4,1)$ gauge. Furthermore, it is also shown that these constraints can be consistently implemented via a variational principle in the Lagrangian.

In total, the condensate that forms reduces the action to
\beq
S_{\rm reduced}=\frac{\sigma\langle\bar{\psi}_i\star \psi_i \rangle}{4}\int_M \star F\,F - \frac{\sigma\langle\bar{\psi}_i\psi_i \rangle}{4}\int_M F\,F +S_{\rm matter}\,.
\eeq
Defining the constants
\beq
\frac{\theta}{8\pi^2} = \frac{\sigma \langle\bar{\psi}_i \psi_i \rangle}{4} \quad \quad \frac{1}{\kappa}=\frac{\sigma\langle\bar{\psi}_i\star \psi_i \rangle}{4}
\eeq
the action reduces to the MacDowell--Mansouri action with a topological $\theta$-term similar to the analogous term in QCD:
\beqa
S_{\rm reduced}&=& \frac{1}{\kappa}\int_M \star F\,F -\frac{\theta}{8\pi^2} \int_M F\,F +S_{\rm matter}\,.
\eeqa
Since the local gauge symmetry has been reduced to $Spin(3,1)$, we can now rewrite the total action in terms of its constituents $\omega$, and $e$ (separating the topological from the non-topological terms, and defining, $\kappa=-\frac{2k}{\ell^2}$ and $k=8\pi G$):
\beqa
S_{\rm reduced}&=& S_{\rm matter}+\frac{1}{k}\int_M \left[\star\, e\,e\,R -\frac{1}{2\ell^2} \star e\,e\,e\,e\right] \nn\\
& & + \int_M\left[-\frac{\ell^2}{2k}\star R\,R -\frac{\theta}{8\pi^2} \,R\,R +\frac{ \theta}{8\pi^2\ell^2}\left(T\,T +2\, e\,e\,R\right)\right]\,.
\eeqa
We now recognize the last term in the second line as the Holst term modifying the Einstein--Cartan action. It appears in combination with the torsion-torsion term reconstructing the so-called Nieh--Yan invariant, which, in the extended formulation presented in \cite{Mer09p} (see also \cite{DatKauSen09,Mer06,Mer08}) and described in Section \ref{Nieh--Yan}, multiplies the BI parameter. If, in fact, we identify $\frac{\theta}{8\pi^2}=\frac{\ell^2}{2 k \gamma}$, the total action becomes:
\beqa
S_{\rm reduced}= S_{\rm matter}&+&\frac{1}{k}\int_M\left[\star\,e\,e\,R -\frac{\Lambda}{6} \star e\,e\,e\,e +\frac{1}{\gamma}\left(e\,e\,R+\frac{1}{2}\,T\,T\right) \right] \nn\\
&-&\frac{\ell^2}{2 k}\int_M\left[ \star \,R\,R +\frac{1}{\gamma} R\,R\right]
\eeqa

Thus, we see that there is a clear sense in which the BI parameter naturally arises from this symmetry breaking mechanism, albeit in conjunction with other couplings. We will show later that this is a generic feature of symmetry breaking mechanisms of this sort, and we will show how these extra couplings can be interpreted. The key feature of the mechanism is the emergence of a parity violating term proportional to the second Chern class of the de Sitter connection, from which one can extract the Nieh--Yan class, which in turn contains the BI term. Later, we will show how the Nieh--Yan class naturally generalizes the Holst term to matter contribution \cite{Mer09p,Mer06,Mer08} and how it leads to the Ashtekar--Barbero constraints of GR \cite{Ash86-87a,Ash86-87b,Bar95a, Bar95b} with fermions.

\section{The degenerate vacuum}
We will now argue that the key features of the model described in the previous section are generic features that one should expect from any theory with exact local de Sitter symmetry, that has de Sitter space (or a state locally isomorphic to de Sitter space) as a ground state. Due to the structure of the gauge group, the de Sitter vacuum defined by $F_\cA$ is degenerate. The emergence of the parameter $\theta$ can be associated with the property of quantum mechanical tunneling between distinguished degenerate vacua of the theory.

Let us begin with a gauge theory based on the connection $\cA$ as the connection in a local patch of a principle $G$-bundle, where $G=Spin(4,1)$, namely the de Sitter group. Take the full set of gauge symmetries to be $Spin(4,1)_M \rtimes Diff_4(M)$, where $Diff_4(M)$ is the group of diffeomorphisms of the base manifold and $Spin(4,1)_M$ is the set of (vertical) gauge transformation of the bundle with base manifold $M$. In other words, $Spin(4,1)_M$ is the set of gauge transformations generated by $g:M\rightarrow Spin(4,1)$. Assume also that the theory has a class of preferred ground states that satisfies $F_{\cA}=0$. We take this theory to define the more fundamental theory, that must reduce to Einstein--Cartan gravity with a positive cosmological constant upon some symmetry breaking mechanism sending the gauge group from $Spin(4,1)_M \rtimes Diff_4(M)$ to $Spin(3,1)_M \rtimes Diff_4(M)$, the latter being the gauge group of Einstein--Cartan gravity. 

If the symmetry breaking mechanism is truly dynamical, the symmetry ``broken'' phase of the theory will still retain the full symmetry of the original theory, but this symmetry will be realized in a non-obvious, typically non-linear manner. Furthermore, it is common that the realization of a dynamic symmetry breaking mechanism requires the introduction of new dynamical fields that serve as order parameters for the mechanism, as in the Stelle-West model where the five dimensional vector field serves as the order parameter to break the symmetry. 
Without direct knowledge of a true dynamical mechanism, the situation can be modeled by an explicit symmetry breaking. In this case, the symmetries of the two theories will be genuinely different. Generically, the existence of homotopically distinct gauge transformations of the theory with larger symmetry, can imply the existence of degenerate vacuum solutions to the symmetry reduced theory. For example, suppose both theories contain solutions of the form $F_A=0$, taken to be the ground state sector. Consider a gauge transformation $g(x)\in Spin(4,1)_M$ that is not continuously deformable to the identity and is not in the subgroup $Spin(3,1)_M$. Apply this gauge transformation to a flat connection $\cA$ to obtain ${}^g\!\cA=g\cA \,g^{-1}-dg g^{-1}$. This connection is still flat since $F_{{}^g\!\cA}=g F_{\cA} \,g^{-1}=0$. This is clearly just a $Spin(4,1)$ gauge transformation, so with respect to the symmetries of the theory with larger symmetry, these solutions should be considered physically equivalent. However, with respect to the the symmetries of the reduced theory, these solutions may be physically distinct, since only the flat sector defined by $F_\cA=0$ possesses this symmetry, and not the full phase space. 

Let us consider two Cauchy slices and the two connections $\cA$ and ${}^g\!\cA$, defined respectively on the two slices. These are connected by {\it continuous} paths which cannot be a gauge orbit, even in the full theory, since the data on the two slices are related by a gauge transformation that is not connected to the identity. Such transitions, between two topologically distinct field configurations are known as instantons. Generically there can be classical solutions representing instantons --- however, even if there are not, these paths will contribute to the path-integral sum. We will now show how this applies to the case at hand. For the purposes of this paper, it is sufficient to restrict attention to spacetime configurations that are flat in the de Sitter sense ($F_{\cA}=0$) in the asymptotic past and asymptotic future. We stress that this is a pragmatic restriction only, intended to simplify the discussion. In particular, this ensures that integrals such as $\int_M F_\cA\,F_\cA$ are true topological invariants even when the boundary of the manifold is taken to be $\partial M=\Sigma_{-\infty} \cup \Sigma_\infty$, the spacelike hypersurfaces in the asymptotic past and future. One could potentially lift this constraint by weaker boundary conditions, or by adding boundary terms to the action, and to the topological integrals (see for example \cite{Eguchi:1980jx}). However, these terms will generically alter the symplectic form. Since we are focusing on a direct relation between this framework and the ordinary canonical framework of the Holst action, we choose to work with the stronger asymptotic boundary conditions for conceptual clarity.

To see this, we first need to characterize the relevant topological properties of the de Sitter group. We will take the base manifold to have the topology of the de Sitter ground state, namely $M=\mathbb{R}\times \mathbb{S}^3$, so that a typical constant time Cauchy slice has the 3-sphere topology. We now wish to consider the set of topologically distinct gauge transformations of the initial data on a Cauchy slice generated by $g:\mathbb{S}^3\rightarrow Spin(4,1)$. Equivalently we could restrict ourselves to time-independent gauge transformations and restrict attention to one spatial slice. This procedure was carried out recently by AR, and the set of homotopically distinct gauge transformations (with respect to the third homotopy group) was constructed explicitly. To summarize, the relevant topological properties of the gauge group are contained in the maximal compact subgroup, which in this case is $Spin(4)$. The relevant topological information that we will need is characterized by the third homotopy group of the gauge group viewed as a manifold, which in our case is given by $\pi_3(Spin(4,1))=\pi_3(Spin(4))=\pi_3(SU(2)\times SU(2)) =\mathbb{Z}\oplus\mathbb{Z}$. Thus, the homotopically distinct gauge transformations are labeled by two integers, and we will denote a class of such generators by $\ou{m}{g}{n}$. It is worth recalling that de Sitter space satisfies the condition $F_\cA=0$ --- thus, applying the gauge transformation to $\cA$ yields an infinite class of flat connections labeled by two integers. More specifically the gauge transformation can be applied explicitly to yield $\ou{m}{\cA}{n}= \ou{m}{g}{n}\cA \ou{m}{g}{n}{}^{-1}-d\ou{m}{g}{n} \,\ou{m}{g}{n}{}^{-1}$. In the full theory these are just gauge transformations, however, in the Einstein--Cartan theory one can proceed to extract the new tetrad, $\ou{m}{e}{n}$, and the new metric $\ou{m}{\mathfrak{g}}{n}$ which turns out to depend only on the difference $q=m-n$. These solutions are physically distinct with respect to $Spin(3,1)_M\rtimes Diff_4(M)$.

Now, consider two connections obtained in this way, $\ou{m_i}{\cA}{n_i}$ and $\ou{m_f}{\cA}{n_f}$, and pull-back the first connection to a $t=-\infty$ slice and the second connection to a $t=\infty$ slice. A continuous path connecting the two configurations can easily be constructed. For example suppose $f(t)$ is a continuous and differentiable function such that $f(-\infty)=0$, $f(\infty)=1$, and $df=0$ at $t=-\infty$ and $t=\infty$. A continuous path is given by
\beq
\cA_*=\ou{m_i}{\cA}{n_i}+f(t)\left(\ou{m_f}{\cA}{n_f}-\ou{m_i}{\cA}{n_i}\right)\,. \label{TypicalPath}
\eeq
This connection is asymptotically flat in the past and future, but in the interior generically the reduced curvature $R_{\omega}-\frac{1}{\ell^2}\left( e\,e\right)$ and the torsion $D_{\omega}e$ are non-zero. To see this we note that (see \cite{RandonodSSpaces} for an explicit demonstration of this)
\beqa
\frac{1}{8\pi^2}\int_M F_{\cA_*}\,F_{\cA_*}= Y_{CS}[\ou{m_f}{\cA}{n_f}(\infty)]- Y_{CS}[\ou{m_i}{\cA}{n_i}(-\infty)]=\Delta m+\Delta n
\eeqa
where $\Delta m=m_f-m_i$ and $\Delta n=n_f-n_i$. As mentioned previously, generically this solution will not be a solution to the Einstein--Cartan field equations, nor is it necessarily a solution to the field equations of the $Spin(4,1)$ theory. However, it {\it will} contribute to the path integral in a calculation of the quantum mechanical tunneling amplitude between the corresponding quantum ground states. 

To see this, let $|\psi_{dS}\rangle=|0,0\rangle$ be the quantum ground state corresponding to de Sitter space. Denote the unitary action of a large gauge transformation on the state by $\ou{m}{G}{n}$ so that one can build a tower of states labelled by two integers, $|m,n\rangle=\ou{m}{G}{n}|0,0\rangle$, which are the quantum analogues of the flat field configurations $\ou{m}{\cA}{n}$. Now consider the physical inner product of two states $|m_f,n_f\rangle $ and $|m_i,n_i\rangle$. Being gauge invariant, the inner product satisfies
\beqa
\langle m_f,n_f\mid m_i,n_i\rangle &=& \langle m_i+\Delta m,n_i+\Delta n\mid m_i,n_i\rangle \nn\\
&=& \langle \Delta m,\Delta n\mid \ou{m_i}{G}{n_i}{}^{-1}\, \ou{m_i}{G}{n_i}\mid 0,0\rangle \nn\\
&=& \langle \Delta m,\Delta n\mid 0,0\rangle
\eeqa
demonstrating that the inner product depends only on the change in $m$ and $n$. The physical inner product can be related to the path integral by
\beqa
\langle m_f,n_f\mid m_i,n_i\rangle=\langle \Delta m,\Delta n\mid 0,0\rangle =\int [\mathcal{D}\Phi\mathcal{D}\cA]_{\{\Delta m,\Delta n\}}\ e^{i S[\cA, \Phi]} \label{InnerProduct}
\eeqa
where $\Phi$ represents the auxiliary matter fields of the $Spin(4,1)$ theory. The measure is restricted to the set of de Sitter connections that are asymptotically flat in the past and future, and connect homotopy sectors that differ by $\Delta m$ and $\Delta n$. A typical path that contributes to the sum in the functional integral is given by (\ref{TypicalPath}). Generically this path-integral will be non-zero giving rise to a quantum mechanical tunneling amplitude between the degenerate vacua. Since there is no known general principle that selects one $|m,n\rangle$ vacuum state over another, we must turn to superpositions of $|m,n\rangle$ states that are stable against quantum mechanical tunneling.\footnote{In general, a superposition of states invariant under the large gauge sector of the theory can be easily introduced at the price of modifying the momentum operator, in the canonical formulation of quantum mechanics. This is reflected in the Lagrangian in the form of new topological terms that typically violate parity. It is worth stressing that in Yang--Mills gauge theories this topological modification can be reabsorbed through a suitable chiral rotation, which produces an analogous term in the fermionic measure, eventually modifying the mass term. Interestingly enough, in the case that at least one of the fermions is massless, we can still argue for the existence of $\theta$-states, but these states will effectively be physically equivalent.} 
This is where the $\theta$-angles enter into the picture. A $\theta$--state is constructed to be stable against quantum mechanical tunneling. Since $\pi_3(Spin(4,1))=\mathbb{Z}\oplus\mathbb{Z}$, there is a two-fold large gauge symmetry, giving rise to the two integers $|m,n\rangle$. Likewise, associated with both large gauge degrees of freedom, there will be independent instanton angles, denoted $\theta_1$ and $\theta_2$. A $\theta$ state is a coherent superposition of the $|m,n\rangle$ constructed to be an eigenstate of the unitary large gauge transformation operator $\ou{-m}{G}{-n}=\ou{m}{G}{n}{}^{\dagger}$ whose eigenvalues are related to the $\theta$-angle by
\beqa
\ou{m}{G}{n}{}^{\dagger}\,|\theta_1,\theta_2\rangle=e^{i\theta_1m} e^{i\theta_2n} \,|\theta_1,\theta_2\rangle\,.
\eeqa
Similar to the coherent states of quantum optics, which are eigenstates of the annihilation operator, the state are eigenstates of the analogous lowering operator: defining $G_{\ua}=\ou{1}{G}{0}$, and $G_{\da}=\ou{0}{G}{1}$, we have $G_\ua^{\dagger}|\theta_1,\theta_2\rangle=e^{i\theta_1}|\theta_1,\theta_2\rangle$ and $G_\da^{\dagger}|\theta_1,\theta_2\rangle=e^{i\theta_2}|\theta_1,\theta_2\rangle$. Noting that $\ou{m}{G}{n}={G_{\ua}}^m {G_{\da}}^n$, it can easily be seen that the above properties are satisfied by
\beq
|\theta_1,\theta_2\rangle= \sum^{\infty}_{m'=-\infty}\sum^{\infty}_{n'=-\infty} e^{i\theta_1m'} e^{i\theta_2n'}|m',n'\rangle\,.
\eeq
Consider now the physical norm of two $\theta$-states, $\langle\theta_1,\theta_2 \mid \theta_1,\theta_2 \rangle$. Using the inner product (\ref{InnerProduct}), this is given by
\beqa
\langle\theta_1,\theta_2 \mid \theta_1,\theta_2 \rangle &=& \sum_{\{m_f,n_f\}}\sum_{\{m_i,n_i\}} \langle m_f,n_f \mid e^{-i \theta_1 m_f}e^{-i\theta_2 n_f}\,e^{i \theta_1 m_i}e^{i\theta_2 n_i} \mid m_i, n_i\rangle \nn\\
&=& \sum_{\{\Delta m,\Delta n\}}\sum_{\{m_i,n_i\}} e^{-i\theta_1 \Delta m}e^{-i \theta_2 \Delta n}\langle m_i+\Delta m,n_i+\Delta n \mid m_i, n_i\rangle \nn\\
&=& \sum_{\{\Delta m,\Delta n\}} e^{-i\theta_1 \Delta m}e^{-i \theta_2 \Delta n}\int \{\mathcal{D}\Phi \mathcal{D}\cA\}_{\{\Delta m,\Delta n\}} \ e^{iS[\cA,\Phi]}\,.
\eeqa
It is often the case that the integer winding numbers can be related to a four dimensional topological integral. Let us suppose that there are two topological functionals $S_{\ua}[\cA, \Phi]$ and $S_{\da}[\cA, \Phi]$, that yield the respective instanton numbers $\Delta m$ and $\Delta n$ when restricted to a set of solutions that tend towards $\ou{m+\Delta m}{\cA}{n+\Delta n} $ at $t=\infty$ and $\ou{m}{\cA}{n}$ at $t=-\infty$. The inner product can then be written:
\beq
\langle\theta_1,\theta_2 \mid \theta_1,\theta_2 \rangle=\int \{\mathcal{D}\Phi \mathcal{D}\cA\}_{all\ \{\Delta m,\Delta n\}} \ e^{-i\theta (S_{\ua}+S_{\da})} \,e^{-i\widetilde{\theta}(S_{\ua}-S_{\da})}\,e^{iS}
\eeq
where it is now understood that the functional integral is to be defined over all $\{m,n\}$ sectors, and we have defined:
\beq
\theta=\frac{1}{2}(\theta_1+\theta_2) \quad \quad \quad \widetilde{\theta}=\frac{1}{2}(\theta_1-\theta_2) \,.
\eeq

The topological functional integrals are defined such that when restricted to the $\{\Delta m,\Delta n\}$ sectors, they yield $S_{\ua} +S_{\da}=\Delta(m+n)$ and $S_{\ua} -S_{\da}=\Delta(m-n)$. One such functional is known, and it was described above: the first topological functional can be identified with the second Chern-class of the de Sitter connection:
\beq
S_{\ua}+S_{\da}=\frac{1}{8\pi^2}\int_M F_{\cA}\, F_{\cA}
\eeq
since this functional is topological and gives the expected winding number. On the other hand, although the normalized three-volume of the throat of the de Sitter solutions can be related to the integer $q$ \cite{RandonodSSpaces}, at this time we do not know of any {\it four}-dimensional topological invariant that can be identified with $S_{\ua}-S_{\da}$. Thus for the remainder of this argument, we will set $\widetilde{\theta}=0$. We would like to stress that this is a pragmatic restriction only: the $\widetilde{\theta}$ term may, indeed,  be physically relevant, and may yield interesting new physical predictions. However, without knowledge of a closed form for the topological functional $S_{\ua}-S_{\da}$, we would be stuck with an additional ambiguity. It should also be mentioned that an alternative approach would be to simply restrict attention to the $q=m-n=0$ sectors. This would have the advantage that at the classical level, the induced metric obtained by extracting the tetrad from $\ou{m}{\cA}{m}$, and computing the metric $\ou{m}{\mathfrak{g}}{m}$, is identically de Sitter space for all $m$ (this procedure has been carried out in \cite{RandonodSSpaces}). Thus, the physical interpretation of the state $|\theta,q=0\rangle$ would be more straightforward. However, there is a technical detail that makes this choice difficult to carry out -- the integers $q=m-n$ and $p=m+n$ are not entirely independent since they must be either both even or both odd. Thus, fixing $\Delta q=0$ in the path integral would require restricting the paths and the measure, $\{\mathcal{D}\Phi \mathcal{D}\cA\}_{\Delta p}$, only to sectors where $\Delta p$ is even. To avoid this difficulty, we will work with in the $\widetilde{\theta}=0$ sector, and simply define the state $|\theta \rangle \equiv |\theta_1,\theta_2\rangle \big|_{\theta_1=\theta_2}$. 

In this case, the total inner product becomes
\beqa
\langle \theta \,|\, \theta \rangle=\int \mathcal{D}\Phi \mathcal{D}\cA \ e^{-i\frac{\theta}{8\pi^2} \int_M F_\cA\,F_\cA} \,e^{iS[\cA,\Phi]}
\eeqa
where it is understood that the path integral is over all $\{\Delta m, \Delta n\}$ sectors. The total effective action is then
\beq
S_{total}=S[\cA, \Phi]-\frac{\theta}{8\pi^2}\int_M F_\cA\,F_\cA\,.
\eeq
Upon symmetry breaking, the base action must reduce to some variant of the Einstein-Cartan action (by assumption), and the second Chern-class can be split into its constituents as in the toy model above. Let us assume that the base action reduces to the Macdowell-Mansouri action with the Dirac action as matter fields. Thus, the total reduced symmetry action becomes:
\beqa
S_{\rm total} &=& S_{Dirac}+\frac{1}{k}\int_M \left[\star e\,e\,R -\frac{1}{2\ell^2} \star e\,e\,e\,e\right] \nn\\
& + &\int_M\left[-\frac{\ell^2}{2k}\star R\,R -\frac{\theta}{8\pi^2} \,R\,R +\frac{ \theta}{8\pi^2\ell^2} \left(T\,T +2\,e\,e\,R\right)\right] \label{ReducedAction1}
\eeqa
where the Dirac action is
\beq
S_{Dirac}=\frac{2i}{3}\int_M \bar{\psi} \star e\,e\,e\,D_{\omega}\psi +D_{\omega}\bar{\psi} \star e\,e\,e\,\psi\,.
\eeq

\subsection{Regaining the Holst action}
We would now like to show that this action reduces to the Holst modification of the Einstein-Cartan action, with some extra topological terms and a non-minimal fermionic coupling. The additional topological terms do not affect the equations of motion, but they do complicate the analysis. Specifically, they modify the definition of the momenta conjugated to the Ashtekar--Barbero variables only, without affecting the identification of the Barbero--Immirzi parameter with the constant coefficient in front of the Nieh--Yan term. So, for the purposes of illustration, to begin with let us first consider the limit when the cosmological constant goes to zero. In taking such a limit, one must first decide on what quantities to fix. To retain the information contained in the parity violating term, we demand that in the limit that the cosmological constant goes to zero, the ratio of the parity even coupling constant, call it $\alpha_+$, and the parity odd coupling constant, $\alpha_-$, must be held fixed. This will ensure that the information about the instanton configurations is preserved in taking the limit. Thus, in taking the limit as $\Lambda\rightarrow 0$ we will fix the ratio:
\beq
\frac{\alpha_+}{\alpha_-}=-\frac{\left(1/\kappa\right)}{\left(\theta/8\pi^2\right)}\,.
\eeq
It is this ratio that can be identified with the ordinary Barbero--Immirzi parameter when the cosmological constant is zero.

To see this, we rewrite (\ref{ReducedAction1}) action as follows:
\beqa
S_{\rm total} &=& \alpha_+ \int_M \star \,F\,F +\alpha_- \int_M F\,F +S_{Dirac}\nn\\
&=& \alpha_+\int_M \left[\star \,F\,F +\frac{\alpha_-}{\alpha_+} \,F\,F \right] +S_{\rm Dirac} \nn\\
&=& -\frac{\ell^2}{2k} \int_M \left[\star \,R\,R + \frac{\alpha_-}{\alpha_+} R\,R\right] +\frac{1}{k}\int_M  \left[\star \,e\,e\,R -\frac{1}{2\ell^2}\star e\,e\,e\,e \right] \nn\\
& & +\frac{1}{k}\int_M \frac{\alpha_-}{\alpha_+}\left[e\,e\,R +\frac{1}{2}\,T\,T \right] +S_{\rm Dirac} \label{ReducedAction2}
\eeqa

Generically, the limit $\ell\rightarrow \infty$ will not converge unless we have $\int\left(\star R\,R+\frac{\alpha_-}{\alpha_+}R\,R\right)=0$. Since this is a topological constraint only, and we have already extracted the relevant topological information from the action, for the purposes of this argument the constraint is not too restrictive. The limit then becomes
\beqa
S_{\rm total} &=& \frac{1}{k}\int_M \left[\star\, e\,e\,R  +\frac{\alpha_-}{\alpha_+}\left(e\,e\,R+\frac{1}{2}\,T\,T\right)\right]\nn\\
& & +\frac{2i}{3}\int_M \bar{\psi} \star e\,e\,e\,D_{\omega}\psi +D_{\omega}\bar{\psi} \star e\,e\,e\,\psi \label{ReducedAction3}
\eeqa
The Barbero--Immirzi parameter can then be viewed as the coupling constant of the topological Nieh--Yan form, $Tr(2\,e\,e\,R+T\,T)=Tr(d(e\,T))$. Although this appears to be distinct from the ordinary Barbero--Immirzi term, when the canonical analysis is carried out explicitly, the resulting constraints of this more general theory are exactly those of the ordinary Ashtekar--Barbero canonical formulation of GR \cite{DatKauSen09}, as we will demonstrate in Section \ref{Nieh--Yan}.

\subsection{The anti-de Sitter group}
In the analysis above we focused on the de Sitter group because it is observationally preferred. However, the main features of the analysis appear to carry through even when the gauge group is taken to be the anti-de Sitter group $Spin(3,2)$ with some important qualifications. The most important distinction involves the topological structure of the group and the topology of AdS space itself. The universal cover of $Spin(3,2)/Spin(3,1)$ is identified with $\text{AdS}_4$, which has topology $\mathbb{R}^4$. A typical spatial slice will then have spatial topology $\mathbb{R}^3$. Thus, to identify the interesting topological features of $Spin(3,2)_M$, one can adopt the ordinary procedure of adding boundary terms to the action, fixing the gauge at spatial infinity, and compactifying the spatial slice to $\mathbb{S}^3$ by adding the point at infinity to the topological manifold. The remaining symmetries at spatial infinity will become true symmetries and the boundary terms in the action will be related to conserved quantities in the bulk. 

Following this procedure one can then analyze the topological structure of $Spin(3,2)_M$. The most important distinction is that whereas $\pi_3(Spin(4,1))=\mathbb{Z}\oplus\mathbb{Z}$, for the anti-de Sitter group we have $\pi_3(Spin(3,2))=\mathbb{Z}$. Furthermore, the generating subgroup of the topologically non-trivial maps $g:\mathbb{S}^3\rightarrow Spin(3,2)$ can be identified as living in a rotation subgroup of the full group. This means that one can still carry out the procedure of defining the AdS connection $\mathcal{A}$, transforming it by a large gauge transformation, and extracting the new tetrad -- however, the new connection will be labeled by a single integer, and the metric that one obtains from the extracted tetrad will be identical to ordinary anti-de Sitter space. Instanton solutions can be constructed in the same way, however, they will be topological solutions connecting the AdS vacuum at past infinity to the same AdS vacuum at future infinity. Nevertheless, the instanton number can still be identified with $\frac{1}{8\pi^2}\int F_\mathcal{A}\,F_\mathcal{A}$, so the $\theta$-states can be constructed and the remainder of the calculation proceeds analogously to the de Sitter case. Again, the Barbero--Immirzi parameter can be identified as a combination of Newton's constant, the cosmological constant, and the $\theta$-parameter.

\section{Canonical analysis}\label{Nieh--Yan}

This section is entirely devoted to the canonical formulation of the gravitational theory modified by the Nieh--Yan invariant. The focus will be on identifying the ratio $\alpha_+/\alpha_-$ with the Barbero--Immirzi parameter. First, we demonstrate that, in fact, the canonical analysis of the modified action containing the Nieh--Yan invariant leads to the Ashtekar--Barbero constraints of pure GR. Second, we demonstrate that, in the presence of spinor matter, the symplectic structure of gravity is not affected, while the torsion-torsion term modifies the fermionic symplectic structure being equivalent to a non-minimal coupling previously introduced \cite{Ran05,Mer06,Mer06p}. The key feature that we wish to highlight, that the two actions are equivalent at the classical, canonical level if the $\int T\,T$ terms in the Nieh-Yan formulation are replaced with the appropriate non-minimal coupling terms of the form $\int \bar{\psi} \,e\,e\,e \,D\psi -D\bar{\psi}\, e\,e\,e \,\psi$ in the ordinary Holst formulation. Since the Hamiltonian framework is the starting point of canonical quantization, the two actions will be equivalent at the quantum level as well.

In order to be as clear as possible, in this section we reintroduce space-time indexes  $\mu,\nu,\rho\dots$ and internal indexes, $I,J,K\dots$, for Lorentz valued tensors. It will be convenient here to use the external dual $*$ as opposed to the internal dual $\star$. The Nieh--Yan modified action reads
\beqa
S_{\rm total}&=&\frac{1}{2k}\,\int_M \left[*\left(e_I\wedge e_J\right)\wedge R^{IJ}  -\frac{\alpha_-}{\alpha_+}\left(e_I\wedge e_J\wedge R^{IJ}-T^I\wedge T_I\right)\right]
\nn \\
&=&\frac{1}{2k}\,\int_M d^4x \det\!^{(4)}\! e\,\left[e^{\mu}_{\ I}e^{\nu}_{\ J} R_{\mu\nu}^{\ \ IJ}+\frac{1}{2\g}\,\epsilon^{IJ}_{\ \ KL}\,e^{\mu}_{\ I}e^{\nu}_{\ J}R_{\mu\nu}^{\ \ KL}-\frac{1}{4\g}\,\epsilon^{\mu\nu\rho\sigma}D_{\mu}e^{\ I}_{\nu}D_{\rho}e^{\ J}_{\sigma}\right]. \label{ReducedAction4}
\eeqa
As mentioned previously, the focus of this Section is on clarifying the role of the constant ratio $\alpha_+/\alpha_-$, which, at the end, will be identified with the Barbero--Immirzi parameter, so, for convenience we redefine it as $\g=\alpha_+/\alpha_-$. Note that the torsion-torsion term has been rewritten in terms of the tetrad fields and covariant derivatives, according to the following definitions: $T^I=\frac{1}{2}\,T^I_{J K}e^J\wedge e^{K}$ with $T^I_{\mu\nu}=D_{[\mu}e^{\ I}_{\nu]}$. The usual expression of the volume 4-form, $e^I\wedge e^J\wedge e^K\wedge e^L=-\epsilon^{IJKL}\det(e)\,d^4x$, with $\epsilon_{0123}=-\epsilon^{0123}=1$, has been used. 

Now, assuming that the space-time $(M,g_{\mu\nu})$ is globally hyperbolic, then, according to Geroch theorem \cite{ger70}, we can define a global time function $t$ in such a way that each surface of constant $t$ is a Cauchy surface $\Sigma^3$ and the space-time topology is $M=\mathbb{R}\times\Sigma^3$. On each surface, the 4-dim metric $g_{\mu\nu}$ induces a Riemannian metric $h_{\mu\nu}$ defined by the first fundamental form, i.e.
\be
h_{\mu\nu}=g_{\mu\nu}+n_{\mu}n_{\nu}\,,
\ee
where $n^{\mu}$ is the normal vector to $\Sigma^3$. Let $t^{\mu}=t^{\mu}(y)$ be a vector field in $M\ni y$, satisfying $t^\mu\nabla_\mu t=t^\mu\p_\mu t=1$, or, equivalently, $t_{\mu}dx^{\mu}=dt$.\footnote{We remark that neither $t$ nor $t^{\mu}$ can be interpreted in terms of physical measurements of time, since one does not know the metric, which is, in fact, the unknown dynamical field in the Einstein theory of gravitation.} The ``time flow'' vector field, $t^{\mu}$, generates a one-parameter group of diffeomorphisms, known as embedding diffeomorphisms, $\phi_t:\mathbb{R}\times\Sigma^3\rightarrow M$, defined as $y(t, {\bf x})\equiv y_t({\bf x})$. This allows us to represent space-time as a one-parameter family of 3-dimensional Cauchy surfaces $\Sigma^3_t$, obtained by smoothly deforming the initial surface $\Sigma^{3}$ and described by the parametric equations $y^{\mu}_t=y^{\mu}_t({\bf x})$, where $t$ denotes surfaces at different ``times''. A general and convenient parametrization of the embedding can be obtained by introducing the normal and tangential components of the deformation vector $t^{\mu}(y)$ with respect to $\Sigma^3$. Namely, we define
\beqa
N\equiv g_{\mu\nu}t^{\mu}n^{\nu}\,,\qquad N^{\mu}\equiv h^{\mu}_{\ \nu}t^{\nu}\,,
\eeqa
respectively called \emph{lapse function} and \emph{shift vector}. As a consequence we have
\beqa
t^{\mu}(t, {\bf x})&=& \left.\frac{\partial y^{\mu}(t, {\bf x})}{\partial t}\right|_{y(t, {\bf x}) =y_t({\bf x})}
=N(t, {\bf x})\,n^{\mu}(t, {\bf x})+N^{\mu}(t, {\bf x})\,.\label{def}
\eeqa
By acting with a Wigner boost on the local basis, we can rotate the zeroth component of the basis in such a way that it results to be parallel, in each point of $\Sigma^3$, to the normal vector $n_{\mu}$, i.e. $n_\mu=e_{\mu}^{\ 0}$. The requirement that this particular choice of the orientation of the local basis be preserved along the evolution fixes the so-called Schwinger, or time gauge; the net result being that the action will no longer depend on the boost parameters. As a consequence, the local symmetry group is reduced from the initial $Spin(3,1)$ to $Spin(3)\simeq SU(2)$, which encodes the remnant spatial rotational symmetry. It can be demonstrated that fixing the time gauge into the action does not affect the consistency of the canonical analysis, this procedure being equivalent to a canonical gauge fixing. After fixing to the time gauge, it is convenient to denote internal spatial indices by lower case $i,j,k,...$ indices. The action (\ref{ReducedAction4}) can be rewritten as follows:
\begin{align}
S_{3+1}=&\int d t d^3x\,\frac{N\det e}{k}\,\left[\frac{t^\mu-N^\mu}{N}\,e_{\ i}^{\alpha}\left(R_{\mu\a}^{\ \ \ 0i}-\frac{1}{2\g}\,\epsilon^{i}_{\ jk}R_{\mu\a}^{\ \ \ jk}\right)\right.
\nn\\
&+\frac{1}{2} e_{\ i}^{\a}e_{\ j}^{\g}\left(R_{\a\g}^{\ \ \ i j}+\frac{1}{\g}\,\epsilon^{ij}_{\ \ 0k}R_{\a\g}^{\ \ \ 0k}\right)
\nn \\
&-\left.\frac{1}{4\g}\,t^{\mu}\epsilon^{\alpha\beta\gamma}D_{\mu}e_{\alpha}^i D_{\beta}e_{\gamma\, i}-\frac{1}{4\g}\,t^{\nu}\epsilon^{\alpha\beta\gamma}D_{\alpha}e^{i}_{\nu}D_{\beta}e_{\gamma\,i}\right]\nonumber
\\
=&\int d t d^3x\det\! ^{3}e\left\{\frac{e_{\ i}^{\a}}{k\g}\left[\mathcal{L}_t\,A^i_{\alpha}-D_\a (t\cdot\omega^{i})+(t\cdot\om^{i}_{\ k})\g K_{\a}^{k}\right]\right.
\nn\\
&-N^{\alpha} \frac{e^{\gamma}_{\ i}}{k\g}\left[F_{\a\g}^{i}-\left(1+\g^2\right)\epsilon^{i}_{\ jk}K^j_{\a}K^k_\g\right]\nn
\\
&-N\frac{\g^2 k}{2}\,\frac{e_{\ i}^{\a}}{k\g}\frac{e_{\ j}^{\g}}{k\g}\left[\epsilon^{i j}_{\ \ k}F^k_{\alpha\gamma}-2\left(\g^2+1\right)K^{i}_{[\a}K^{j}_{\g]}-2\left(\g+\frac{1}{\g}\right)\epsilon^{i j}_{\ \ k}D_{[\a}K^k_{\g]}\right]\nonumber
\\
&-\left.\frac{eN}{4\g}\,\epsilon^{\alpha\beta\gamma}D_{\beta}e_{\gamma i}\left[\mathcal{L}_t\,e^i_{\alpha}+\left(t\cdot\omega_{\ k}^{i}\right)e^{k}_{\alpha}\right]\right\}\,,\label{act2}
\end{align}
where $\epsilon_{ijk}=\epsilon_{0ijk}$, with $\epsilon_{123}=\epsilon^{123}=1$; $t\cdot \omega^i=-\frac{1}{2}\,\epsilon^i_{\ j k}t^\mu \omega_\mu^{jk}$ and $t\cdot \omega^{ik}=t^\mu \omega_\mu^{ik}$. The following definitions have been used:
\begin{subequations}
\begin{align}
A^{i}_{\alpha}&=\g\omega^{0i}_{\alpha}-\frac{1}{2}\,\epsilon^{i}_{\ jk}\omega^{jk}_{\alpha}=\g K^{i}_{\a}+\Gamma^{i}_{\a}\,,
\\
F^{i}_{\a\g}&=2\partial_{[\a}A^{i}_{\g]}+\epsilon^{i}_{\ jk}A^j_{\a}A^k_{\g}\,.
\end{align}
\end{subequations}
Note that the $\mathfrak{so}(3)$ (or equivalently $\mathfrak{su}(2)$) valued 1-form $K^{i}_{\a}$ is related to the extrinsic curvature (or second fundamental form) of the Cauchy surfaces $\Sigma^3_t$ by $K^{i}_{\a}=-h^{\b\g}e_\g^i K_{\a\b}$ (where $K_{\a\b}=h_\a^\mu h_\b^\nu \nabla_{(\mu} n_{\nu)}$), while $\Gamma_{\alpha}^i=-\frac{1}{2}\,\epsilon^{i}_{\ j k}\omega^{jk}_{\a}$ is the $\mathfrak{so}(3)$ valued 3-dim spin connection 1-form. The new variable $A^{i}_\a$ is known as Ashtekar--Barbero connection 1-form and its curvature $F^{i}_{\a\g}$ is related to the Riemann curvature associated to $\Gamma^{i}_{\a}$ by the following formula:
\begin{equation}
F^i_{\a\b}=R^i_{\a\b}+2\g D_{[\a}K^i_{\b]}+\g^2\epsilon^{i}_{\ jk}K^{j}_\a K^k_\b\,.
\end{equation}
At this point we can define the momenta conjugated to the fundamental variables and following the Dirac procedure. For the sake of brevity, it is worth distinguishing among the canonical variables those playing the role of Lagrange multipliers, namely the set $(N,N^\a,t\cdot\omega^i,t\cdot\omega^i_{\ k})$; while $(A^i_\a,\Gamma^j_\b,e^k_\g)$ incorporate the dynamics of the system, which will be limited by the appearance of constraints.\footnote{This distinction is not essential at all at this point of the canonical analysis. One can, in fact, go forward in the procedure treating all the variables as dynamical and then reducing the system, recognizing the Lagrange multipliers after having studied the primary constraints and the dynamics described by the total Hamiltonian. Nevertheless, the mathematical structure of the system is particularly suitable to make this kind of distinction at the very beginning, the advantage being that the procedure becomes shorter and hopefully more clear.}

The momenta conjugated to the dynamical variables are:
\begin{subequations}
\begin{align}
A^i_\a:\,\rightarrow\,P^\a_i=\frac{e}{k\g}\,e^\a_i\,,
\\
e^j_\b:\,\rightarrow\,\pi^\b_j=-\frac{Ne}{4\g}\,\epsilon^{\b\a\g}D_{\a}e_{\g i}\label{e-momentum}\,,
\\
\Gamma^{k}_\g:\,\rightarrow\,\Pi^\g_i=0\label{primary con}\,,
\end{align}
\end{subequations}
and the total Hamiltonian reads:
\begin{align}
H=&\int d^3x \left\{-\left(t\cdot\omega^{i}\right)D_\a P^\a_i-\left(t\cdot\om^{i}_{\ k}\right)\g K_{\a}^{k}P_i^\a\right.
\nn\\
&+N^\a P^\g_i\left[F_{\a\g}^{i}-\left(1+\g^2\right)\epsilon^{i}_{\ jk}K^j_{\a}K^k_\g\right]
\nn\\
&+\frac{N}{2}\sqrt{\frac{\g}{kP}}P_{i}^{\a} P_{j}^{\g}\left[\epsilon^{i j}_{\ \ k}F^k_{\alpha\gamma}-2\left(\g^2+1\right)K^{i}_{[\a}K^{j}_{\g]}-2\left(\g+\frac{1}{\g}\right)\epsilon^{i j}_{\ \ k}D_{[\a}K^k_{\g]}\right]
\nn
\\
&-\left.\left(t\cdot\omega_{\ k}^{i}\right)e^{k}_{\alpha}\pi^{\a}_i+\lambda_{\g}^i\Pi^\g_i\right\}\,,\label{tot hamiltonian}
\end{align}
where $\lambda_{\g}^i$ is a Lagrange multipliers for the primary constraint (\ref{primary con}). Once defined the usual symplectic structure, we can proceed to the calculation of the secondary constraints. In particular, by calculating the Poisson bracket between the total Hamiltonian and the momentum $\Pi^{\g}_i$, conjugate to the 3-dimensional spin connection, we obtain a non-trivial secondary second class constraint. Without entering in the details, we only stress that this second class constraint can be solved and the 3-dimensional spin connection can be expressed as function of the momentum $P^{\a}_i$. In geometrical terms this corresponds to find the solution of the spatial components of 3-torsion, which, as is well known, allows one to express the spin connection as a function of the triads in time gauge. As an immediate consequence the momentum conjugated to the 3-dimensional spin-connection vanishes strongly, as well as the combination $D_{\alpha}P^{\alpha}_i=0$, which is, in fact, a trivial consequence of the compatibility condition. Interestingly enough, the momentum (\ref{e-momentum}) also vanishes strongly as consequence of the solution of the second class constraint, which essentially brings in the information about the vanishing of spatial components of torsion. Now the total Hamiltonian can be drastically simplified. In particular, the dynamics is completely described by the variables $A^i_\a$ and their conjugate momenta $P^\a_i$, which have to satisfy a set of seven first class constraints, namely
\begin{subequations}
\begin{eqnarray}
\mathcal{G}_i &:=& \partial_\a P^\a_i+\epsilon_{i j}^{\ \ k}A_{\a}^{j}P_k^\a\approx 0\,,
\\
\mathcal{H}_{\a} &:=& P^\g_i\left[F_{\a\g}^{i}-\left(1+\g^2\right)\epsilon^{i}_{\ jk}K^j_{\a}K^k_\g\right]\approx 0\,,
\\
\mathcal{H}&:=&\sqrt{\frac{\g}{kP}}P_{i}^{\a} P_{j}^{\g}\left[\epsilon^{i j}_{\ \ k}F^k_{\alpha\gamma}-2\left(\g^2+1\right)K^{i}_{[\a}K^{j}_{\g]}-2\left(\g+\frac{1}{\g}\right)\epsilon^{i j}_{\ \ k}D_{[\a}K^k_{\g]}\right]\approx 0\,,
\end{eqnarray}
\end{subequations}
where in the first weak equation we have used the strong equation $D_{\alpha}P^{\alpha}_i=0$, in order to rewrite the rotational constraint as the typical Gauss constraint of Yang--Mills gauge theories. This set of constraints is equivalent to the following one:
\begin{subequations}
\beqa
\mathcal{G}_i &:=&\partial_\a P^\a_i+\epsilon_{i j}^{\ \ k}A_{\a}^{j}P_k^\a\approx 0\,,
\\
\mathcal{H}_{\a}&:=& P^\g_iF_{\a\g}^{i}\approx 0\,,
\\
\mathcal{H}&:=& \sqrt{\frac{\g}{kP}}P_{i}^{\a} P_{j}^{\g}\left[\epsilon^{i j}_{\ \ k}F^k_{\alpha\gamma}-2\left(\g^2+1\right)K^{i}_{[\a}K^{j}_{\g]}\right]\approx 0\,,
\eeqa
\end{subequations}
which is the usual set of constraints of the Ashtekar--Barbero canonical formulation of General Relativity. 

So, the parameter $\g=\frac{\a_+}{\a_-}$ appearing in front of the of the Nieh--Yan invariant can be identified with the Barbero--Immirzi ambiguity of Loop Quantum Gravity as claimed above.

The naturalness of this identification can be easily motivated also from a different perspective. To this aim, let us reconsider the action (\ref{ReducedAction4})
\begin{equation}
S_{\rm total}=\frac{1}{2k}\,\int_M \left[e_I\wedge e_J\wedge\star R^{IJ}  -\frac{1}{\g}\left(e_I\wedge e_J\wedge R^{IJ}-T^I\wedge T_I\right)\right]\,.
\end{equation}
As is well known, an arbitrary variation of the action can be split in a boundary and a bulk term, respectively $J$ and $B$. The requirement that the bulk term $B$ vanishes provides the classical equations of motion:
\begin{subequations}
\begin{align}
T^I=de^I+\omega^I_{\ J}\wedge e^J&=0\,,
\\
\epsilon_{IJKL}\,e^J\wedge R^{KL}&=0\,,
\end{align}
\end{subequations}
while the (pre)symplectic structure can be obtained by differentiating the boundary term $J$, namely
\begin{align}
J=&-\frac{1}{2k\g}\int_{\partial M} e_I\wedge e_J\wedge\delta\left(\omega^{IJ}-\frac{\g}{2}\,\epsilon^{IJ}_{\ \ KL}\omega^{KL}\right)\nn
\\
&+\frac{1}{k\g}\int T^I\wedge \delta e_I\,.
\end{align}
In particular, on the space of regular solutions of the equations of motion (covariant phase space), the symplectic structure we obtain is 
\begin{align}
\Omega=-\delta J=\frac{1}{k\g}\int_{\Sigma}\delta_{[1}\left(e_I\wedge e_J\right)\wedge\delta_{2]}\left(\omega^{IJ}-\frac{\g}{2}\,\epsilon^{IJ}_{\ \ KL}\omega^{KL}\right)\nn
\\
=-\frac{1}{2k\g}\int_{\Sigma}\delta_{[1}\left[*\left(e_I\wedge e_J\right)\right]\wedge\delta_{2]}\left[\g\omega^{IJ}+\frac{1}{2}\,\epsilon^{IJ}_{\ \ KL}\omega^{KL}\right]\,,
\end{align}
for any two vectors $\delta_1$ and $\delta_2$ tangent to the space of regular solutions. Some general arguments allows one to demonstrate that, in fact, the space of classical solutions as well as the above symplectic structure are independent of the Barbero--Immirzi parameter, $\g$. So, the Barbero--Immirzi parameter does not affect the classical dynamics being the coupling constant of a true topological term, namely the Nieh--Yan density; nevertheless, according to the arguments presented in the previous Sections, it well affects the quantum dynamics, being related to the instanton angle, $\theta$. 

The form of the symplectic structure given above and resulting from the Nieh--Yan modified action is equivalent to the one in \cite{AshLew04}, with the remarkable difference that now the parallel with the $\theta$-angle of Yang--Mills gauge theories, mentioned in \cite{AshLew04} can be completed, providing also an interesting scenario to explain the presence of inequivalent $\g$-sectors in gravity (see also \cite{Mer09p} and \cite{Mer08} to have a complete picture of the role of the Nieh--Yan in relation to the Barbero--Immirzi ambiguity from the the canonical General Relativity perspective).

The last step is the introduction of fermionic matter, which, as is well known, modifies the structure equation producing a non-vanishing torsion contribution. So, let us consider the following action:
\begin{align}
S_{\rm total}&=\frac{1}{2k}\,\int_M \left[e_I\wedge e_J\wedge \star R^{IJ}  -\frac{1}{\g}\left(e_I\wedge e_J\wedge R^{IJ}-T^I\wedge T_I\right)\right]\nn
\\
&+\frac{i}{2}\int_M* e_I\,\left[\bar{\psi}\,\gamma^I\,D\psi+(D\bar{\psi})\,\gamma^I\,\psi\right]\,,
\end{align}
and, as we did before, let us split the boundary from the bulk contribution under a general variation of the action. From the requirement that the bulk term vanishes we obtain the equations of motion:
\begin{subequations}
\begin{align}
T^I=-\frac{k}{4}\,\epsilon^I_{\ JKL}e^J\wedge e^K J_{(A)}^L\,,
\\
e_J\wedge\star R^{IJ}+\frac{ik}{2}*\left(e^I\wedge e_J\right)\wedge\left(\bar{\psi}\,\gamma^J\,D\psi+(D\bar{\psi})\,\gamma^J\,\psi\right)=0\,,
\\
i* e_I\wedge \g^ID\psi=0\,,
\\
-i* e_I\wedge D\bar{\psi}\g^I =0\,.
\end{align}
\end{subequations}
While the boundary term reads:
\begin{align}
J=&-\frac{1}{2k\g}\int_{\partial M} e_I\wedge e_J\wedge\delta\left(\omega^{IJ}-\frac{\g}{2}\,\epsilon^{IJ}_{\ \ KL}\omega^{KL}\right)\nn
\\
&+\frac{1}{k\g}\int T^I\wedge \delta e_I\nn
\\
&-\frac{i}{2}\int_{\partial M}* e_I\left(\delta\bar{\psi}\gamma^I\psi-\bar{\psi}\gamma^I\delta\psi\right)\,.
\end{align}
By varying the boundary term on the space of regular solutions of the equations of motion, we obtain the following symplectic structure:
\begin{align}
\Omega=&-\frac{1}{2k\g}\int_{\Sigma}\delta_{[1}\left[*\left(e_I\wedge e_J\right)\right]\wedge\delta_{2]}\left[\g\omega^{IJ}+\frac{1}{2}\,\epsilon^{IJ}_{\ \ KL}\omega^{KL}\right]\nn
\\
&+\frac{i}{2}\int_{\Sigma}\delta_{[1}\bar{\psi}\delta_{2]}\left[* e_I\left(1-\frac{i}{\g}\g^5\right)\g^I\psi\right]+\delta_{[1}\left[* e_I\bar{\psi}\g^I\left(1-\frac{i}{\g}\g^5\right)\right]\delta_{2]}\psi\,.
\end{align}
In other words, the presence of the Nieh--Yan term does not modify the gravitational symplectic structure, which still corresponds to that of the Ashtekar--Barbero formulation of gravity, but modifies the fermionic symplectic structure in a way completely equivalent to the presence of a non-minimal coupling, $\left(1-\frac{i}{\g}\g^5\right)$, between gravity and matter. This non-minimal coupling, originally introduced elsewhere \cite{Ran05,Mer06,Mer06p}, reduces to the left or right projectors respectively when $\g=\pm\,i$, thus allowing one to easily obtain the well known action for (anti)self-dual gravity coupled to matter \cite{AshRomTat89}. We stress that, starting from the Nieh--Yan modified action, the parameter contained in the non-minimal coupling term naturally corresponds to the Barbero--Immirzi parameter. So, in agreement with \cite{Ale08}, we claim that the most natural choice for the non-minimal coupling constant on the fermionic sector is the Barbero--Immirzi parameter. Moreover, we can say that the most natural starting point is the Nieh--Yan modified action for gravity and minimally coupled fermions, which naturally corresponds to a specific effective non-minimal coupling; whereas any other different choice would spoil the topological interpretation of the Barbero--Immirzi parameter, possibly introducing further ambiguities \cite{Ale08}. 

\section{Subtleties}
In this section we discuss some subtleties in the construction. The emergence of the Nieh-Yan term is the key ingredient that allows for an identification of the Immirzi parameter with a topological term. On the other hand, in our construction, the Nieh-Yan term did not come alone but came in conjunction with the second Chern-class of the spin connection. This term complicated the analysis. In light of this it may seem natural to isolate the Nieh-Yan term alone. For example, since the mechanism relies on a breaking of $Spin(4,1)$ symmetry down to $Spin(3,1)$, it is natural to look for instantons that emerge from group elements in the quotient space $Spin(4,1)/Spin(3,1)$. Since the quotient can be identified with de Sitter space itself, It can be checked that $\pi_3(Spin(4,1)/Spin(3,1))=\pi_3(\mathbb{R}\times \mathbb{S}^3)=\mathbb{Z}$. It seems natural that corresponding index of a field configuration would be encoded in the difference of the second Chern classes of $\cA$ and $\omega$, which can be identified with Nieh-Yan class:
\beq
\int_M F_{\cA}\,F_{\cA} -\int_M R\,R=-\frac{1}{\ell^2}\int_M (T\,T+2\,e\,e\,R)\,. \label{NiehYan}
\eeq
This was the viewpoint adopted in \cite{Zanelli:NiehYan}, where a specific field configuration was given whose index was encoded entirely in the Nieh-Yan functional. Unfortunately, this procedure does not work in general, the precise reasons for which are illuminated in a work in preparation \cite{RandonoWise}. Briefly, the reason why it does not work in general can be traced back to the structure of the group $Spin(4,1)$ viewed as a reductive Cartan algebra (see \cite{Wise:MMgravity}\cite{Wise:2009fu} for a review of this construct). Generically, a Lie algebra is said to be reductive if it admits an orthogonal split $\mathfrak{g}=\mathfrak{h}\oplus \mathfrak{p}$ where $\mathfrak{h}$ is a sub algebra, and $\mathfrak{p}$ is generically just a vector space on which $\mathfrak{h}$ acts . The reductive algebra is, in addition said to be symmetric if there is a grading of the algebra where $\mathfrak{h}$ are the even elements and $\mathfrak{p}$ are the odd elements. In the reductive case the algebra has the generic structure
\beq
[\mathfrak{h},\mathfrak{h}]\subseteq \mathfrak{h} \quad \quad [\mathfrak{h},\mathfrak{p}]\subseteq \mathfrak{p}\,. \label{ReductiveAlgebra}
\eeq
In addition to this, if the algebra is {\it symmetric}, we will have the additional condition
\beq
[\mathfrak{p},\mathfrak{p}]\subseteq \mathfrak{h} \label{SymmetricAlgebra}
\eeq
since the left hand side, being difference of the product of two odd elements, must be even. In general for an arbitrary reductive algebra, one can construct a generalization of the three-dimensional Nieh-Yan functional just as above in (\ref{NiehYan}), where the connection is replaced with $\cA=\omega+\xi$. The $\mathfrak{h}$ part of the curvature $F_{\cA}$ is the corrected curvature $F_{\omega}+\frac{1}{2}[\xi,\xi ]{}_{\mathfrak{h}}$, and the $\mathfrak{p}$ part of the curvature is the corrected torsion $D_\omega \xi+\frac{1}{2}[\xi,\xi ]{}_{\mathfrak{p}}$. If the bundles are trivial, the generalized Nieh-Yan class can be written as a pure boundary term given by
\beqa
\int_M \langle F_{\cA}\,F_{\cA} - F_\omega \,F_\omega \rangle &=& \int_M \langle D_{\om}\xi \,D_{\om}\xi +[\xi,\xi ]_{\mathfrak{h}}\,F_{\om}+ [\xi,\xi ]_{\mathfrak{p}} \, D_{\om}\xi \rangle \nn\\
&=& \int_{\partial M} \langle \xi\,D_{\om}\xi +\frac{2}{3} \xi\,\xi \,\xi \rangle
\eeqa
where $\langle \cdot ,\cdot \rangle$ is an inner product on the algebra with respect to which $\mathfrak{h}$ and $\mathfrak{p}$ are orthogonal. In the case when the algebra is symmetric, we have $\frac{1}{2}[\xi,\xi]_{\mathfrak{p}}=0$ and the inhomogenous part of the Nieh-Yan functional is identically zero:
\beq
\int_{\partial M} \langle \xi\,D_{\om}\xi +\frac{2}{3} \xi\,\xi \,\xi \rangle \stackrel{symmetric}{=} \int_{\partial M} \langle \xi \,D_{\omega}\xi  \rangle\,.
\eeq
However, it is precisely this inhomogenous term is necessary for the functional to pick up the proper index of the field configuration. To see this, define the three-dimesional Nieh-Yan functional on an arbitrary three-manifold $\Sigma$, and evaluate it on a fixed connection $\cA$. Suppose that the connection is flat (and torsion free) connections on $\Sigma$. There full curvature vanishes so $F_{\cA}=0$ and $\tau=0$. The winding number of the large gauge transformation is a quantity that is independent of the field configuration. However, restricted to the set of flat connections, if the algebra is symmetric we have
\beqa
\int_{\partial M} \langle \xi\,D_{\om}\xi +\frac{2}{3} \xi\,\xi \,\xi \rangle=\int_{\partial M} \langle \xi\,D_\omega \xi  \rangle=0\,.
\eeqa
This holds identically. Since a gauge transformation, large or small, takes flat connection to flat connections, the Nieh-Yan functional is identically invariant under a gauge transformations. Thus, the Nieh-Yan functional for symmetric reductive connections does not give the index of the field configuration. Indeed one can check explicitly that for the connection (\ref{TypicalPath}), 
\beq
\int_M F_{\cA}\,F_{\cA} -\int_M R\,R=0.
\eeq
On the other hand the second Chern-class, $\int_M F_{\cA} \,F_{\cA}$, does give the expected result. Thus, we are forced to use this functional, and to deal with the additional topological terms in the action.

\subsection{What is the role of the Nieh-Yan term?}
In light of the above discussion, it is worth commenting on the specific role of the Nieh-Yan term in this picture. Since the Nieh-Yan term alone does not encode the proper index of the gauge field configuration, it may seem superfluous in the construction of the theta sectors. Indeed, the reader may have noticed that the relevant index that gives rise to the Immirzi parameter upon symmetry reduction is the particular combination $p=m+n$. For a group element $\ou{m}{g}{n}\in Spin(4,1)_M$ given by the specific representation presented in \cite{RandonodSSpaces}, it can be shown that when $q=m-n=0$ the resulting group element $\ou{m}{g}{m}$ is in a $Spin(3,1)_M$ subgroup. This is the reason why the metric $\ou{m}{\mathfrak{g}}{n}$ depends only on $q$, whereas the tetrad depends on both $p$ and $q$. In light of this, it may seem that one could carry out the entire construction of the theta-sectors using the second Chern-class of the spin connection alone, obviating the need for the introduction of the Nieh-Yan class and resulting the Immirzi term. What role then does the Nieh-Yan term play in this construction?

The interpretation that we would like to put forward is that the Nieh-Yan term, and hence the Immirzi parameter, is a vestigial artifact left over from the breaking of the $Spin(4,1)$ symmetry to $Spin(3,1)$. To understand this better, we recall that the construction of the large sector of the $Spin(4,1)_M$ gauge group in \cite{RandonodSSpaces} relied on the identification of a maximal compact subgroup via a homeomorphism $G\approx H\times \mathbb{R}^n$, which can be found for any connected, semi-simple, non-compact Lie group $G$. The maximal compact subgroup, $H$, is not unique, but is {\it essentially unique}. This means that it is unique only up to conjugation $H\rightarrow H'=gHg^{-1}$ for an arbitrary element $g\in Spin(4,1)_M$. Although the topological content of $H'$ is then same as $H$, in general the two groups cannot be identified. For example, consider a group element $\ou{m}{g}{n}$ with winding number 
\beq
w(\ou{m}{g}{n})=\frac{1}{24\pi^2}\int_{\mathbb{S}^3} d\ou{m}{g}{n}\,\ou{m}{g}{n}{}^{-1} \w d\ou{m}{g}{n}\,\ou{m}{g}{n}{}^{-1}\w d\ou{m}{g}{n}\,\ou{m}{g}{n}{}^{-1} =m+n\,.
\eeq
Now conjugate $\ou{m}{g}{n}$ by an arbitrary identity connected group element $g\in Spin(4,1)_M$, to define $\ou{m}{h}{n}=g\,\ou{m}{g}{n}\,g^{-1}$. Since $g$ is connected to the identity, the winding number will not change so that, $w(\ou{m}{h}{n})=m+n$. On the other hand, generically $\ou{m}{h}{n}$ will not be in the original maximal compact subgroup $H$.

This has the following practical implication. Suppose we are given a gauge field configuration $\cA_*$ analogous (\ref{TypicalPath}) in the sense that the connection restricted to spatial slice at $t=\infty$ is related to the connection restricted to the $t=-\infty$ spatial slice by a large gauge transformation generated by $\ou{m}{g}{m}$. In this case, the index of the field configuration is given by the second Chern-class of the spin connection alone because of the identity:
\beqa
2\Delta m &=& \frac{1}{8\pi^2} \int_M F_{\cA_*} \,F_{\cA_*} \label{FF}\\
&=& \frac{1}{8\pi^2} \int_M R_{\omega_*} \,R_{\omega_*}+\frac{1}{8\pi^2 \ell^2}\int_{\mathbb{S}^3_{\infty}} e_*\,T_*+-\frac{1}{8\pi^2 \ell^2}\int_{\mathbb{S}^3_{-\infty}} e_*\,T_*\nn\\
&=& \frac{1}{8\pi^2} \int_M R_{\omega_*} \,R_{\omega_*}\,. \label{RR}
\eeqa
The last line follows from the fact that $\ou{m}{g}{m}\in SU(2)_M \subset Spin(3,1)_M$ and the torsion and tetrad transform tensorially (i.e. homogenously) under $Spin(3,1)$. On the other hand, just as in the previous paragraph, one can conjugate the curvature by making a gauge transformation by an arbitrary identity connected element $g\in Spin(4,1)_M$ to define ${}^g\!\cA_*=g\cA_* g^{-1} -dg\,g^{-1}$ and $F_{{}^g\!\cA_*}=g F_{\cA_*} g^{-1}$. This transformation will not effect the winding number since it is gauge invariant, however, generically such a transformation will mix the even and odd components of the curvature, or the reduced curvature and the torsion respectively. Thus, after the gauge transformation, one cannot identify $\int_M F_{{}^g\!\cA_*}\,F_{{}^g\!\cA_*}$ with $\int_M R_{{}^g\omega_*}\,R_{{}^g\omega_*}$ since the new connection ${}^g\!\cA_*$ restricted to the spatial slice at $t=\infty$ is related to the connection ${}^g\!\cA_*$ restricted to the spatial slice at $t=-\infty$ by a gauge transformation generated by $\ou{m}{h}{m}=g \,\ou{m}{g}{m} \,g^{-1}$, but $\ou{m}{h}{m}$ will not necessarily reside in the original $Spin(3,1)_M$ subgroup.

Thus, although the Nieh-Yan term alone does not yield the proper index of the gauge configuration, it is a necessary component of the $Spin(4,1)_M$ topological invariant that does. Thus, we offer that the Nieh-Yan term should be viewed as a vestigial term left over from the breaking of the gauge group $Spin(4,1)_M \rtimes Diff_4(M)$ to $Spin(3,1)_M \rtimes Diff_4(M)$. This suggests that the true meaning and physical consequences of the Immirzi parameter should be properly analyzed in the context of a full de Sitter gauge theory.

\section{Concluding Remarks}
Let us now summarize the proposal we have put forward in this article. The goal of the present work is to describe the Immirzi parameter as a one-parameter quantization ambiguity associated with the topological structure of a gauge theory, in direct analogy with the well known $\theta$-ambiguity of the gauge theories of the standard model. Since the standard construction begins with the Holst modified action, which is the ordinary Einstein-Cartan action with an additional parity violating term $\frac{1}{\g}\int e\,e\,R$ that is not by itself topological, the first step is to reformulate the theory so that the Immirzi parameter becomes the coupling constant of a genuine topological term. We have shown that in the presence of fermions, the Holst modification term can always be replaced with a Nieh-Yan term, with an additional modification of the non-minimal coupling of the fermions. The key point is that the Nieh-Yan term is itself topological. A detailed analysis of the canonical theory reveals that the two formulations are canonically equivalent due to the existence of second class constraints that must be imposed strongly. 

The next step is identify the presence of the Nieh-Yan term as a quantization ambiguity resulting from the topological structure of the gauge theory. Our proposal is to enlarge the gauge group from the local Lorentz group $Spin(3,1)$, to the de Sitter group $Spin(4,1)$. This relies on a well-established formulation of gravity as a gauge theory based on the de Sitter group. This construction, and the topology of the new gauge group allows for an infinite class of flat connections, with de Sitter space being simply one member of the class of classical solutions. Since these classical solutions are simply reflections of the topological structure of the gauge theory, the quantum theory will reflect this structure as well in the form of an infinite class of degenerate ``ground states". From generic arguments, these states will not be true ground state since they are not stable against quantum mechanical tunneling. The true ground state will be a coherent superposition of these states, constructed at the cost of the introduction of a new parameter, $\theta$, into the theory. In turn, this parameter emerges in the effective action as the coupling constant of a topological term, $\frac{\theta}{8\pi^2}\int F_\cA F_\cA$, the second Chern class of the de Sitter connection. 

The relation with the Immirzi parameter and the Holst term comes from the recognition that the Nieh-Yan class is an integral piece of the second Chern class of the de Sitter connection. This can be seen upon symmetry breaking of the gauge theory from the gauge group $Spin(4,1)_M$ to $Spin(3,1)_M$, which isolates the Nieh-Yan term. The Immirzi parameter can then be identified in terms of the cosmological constant, Newton's constant, and the $\theta$-parameter. 

Viewed from this perspective, the Immirzi term (now the Nieh-Yan term) can be interpreted as a vestigial term left over from breaking of the symmetry of a larger gauge theory, and its presence is a reflection of the topological structure of the larger gauge group. Thus, a true understanding of the meaning and ramifications of the Immirzi parameter should come from an analysis of a de Sitter gauge theory. Ultimately, the hope is that such an analysis will shed light on the ubiquitous but mysterious nature of the Immirzi parameter in canonical quantum gravity. 

\section{Acknowledgments}
AR would like to thank Roberto Percacci, Rafael Sorkin, and Laurent Freidel for discussions and comments. In conducting this research, AR was supported by NSF grant OISE0853116. SM was supported in part by NSF grant PHY0854743, The George A. and Margaret M. Downsbrough Endowment and the Eberly research funds of Penn State.

\begin{appendix}
\section{Notation and Conventions}\label{Appendix}
In this section we review the conventions and notation that we have employed in this paper. Throughout we are working with a four-dimensional differential, Riemannian manifold with metric signature $(-,+,+,+)$. To facilitate algebraic manipulation we have employed an index free Clifford algebra notation in this paper. We adopt the complex $4\times 4$-matrix representation of the Clifford algebra defined by $\g^I \g^J+\g^J\g^I=2\eta^{IJ}$. The de Sitter Lie algebra is then spanned by the basis elements $\{ \frac{1}{2}\g^{[I}\g^{J]}\,,\, \frac{1}{2}\g_5 \g^K \}$ and the anti-de Sitter algebra is spanned by $\{ \frac{1}{2} \g^{[I}\g^{J]} \,,\, \frac{1}{2} \g^K \}$. Here $\g_5 =\frac{i}{4!}\epsilon_{IJKL}\g^I\g^J\g^K\g^L=i\g^0\g^1\g^2\g^3$ where $\epsilon_{0123}=-\epsilon^{0123}=1$. We will also use the notation $\star=-i\g_5$ to emphasize the interpretation of this objection as the duality operator on the $\mathfrak{so}(3,1)$ vector spaces. The de Sitter connection $\mathcal{A}$ is (locally) a one-form that takes values in $\mathfrak{spin}(4,1)$ and we identify components by $\mathcal{A}=\omega + \frac{1}{\ell}\g_5 e$ where $\omega=\frac{1}{4} \g_{[I}\g_{J]}\,\omega^{IJ}$ is the spin connection and $e=\frac{1}{2}\g_I \,e^I$ is the tetrad. We will generically denote the curvature with a subscript to identify the connection, as in $F_{\cA}\equiv d\cA+\cA\,\cA$. However, the symbol $R$ will be reserved for the spin connection so that $R\equiv F_\om=d\om+\om\,\om$.

When writing integrals to shorten formulas, we adopt the practice of dropping the explicit $Tr(\cdot,\cdot)$ over the Clifford algebra and dropping explicit wedge products between differential forms. The most relevant trace formulas are listed below. Given the Lie algebra valued differential forms $A=\frac{1}{4}\g_{[I}\g_{J]} \,A^{IJ}$, $B=\frac{1}{4}\g_{[I}\g_{J]} \,B^{IJ}$, $U=\frac{1}{2}\g_I \,U^I$, and $V=\frac{1}{2}\g_I \,V^I$, we have
\beqa 
Tr(A\w B)&=& -\frac{1}{2}\,A^{IJ}\w B_{IJ} \nn\\
Tr(\star \,A\w B) &=& Tr(A\w\star\, B)=\frac{1}{4}\,\epsilon_{IJKL}\,A^{IJ} \w B^{KL} \nn\\
Tr(U \w A) &=& Tr(\star \,U\w A)=0 \nn \\
Tr(U \w V)&=&U_I \w V^I \nn\\
Tr(\star \,U \w V) &=& 0
\eeqa
Using this, the identification between the gravitational action in the fundamental representation and the action in the adjoint representation is
\beq
\frac{1}{k}\int \star\, e\,e\,R +\frac{1}{\gamma}\left(e\,e\,R +\frac{1}{2}\,T\,T\right) = \frac{1}{4k}\int \epsilon_{IJKL}\,e^I \w e^J \w R^{KL} -\frac{2}{\g}\left( e_I \w e_J \w R^{IJ} -T_I \w T^I \right)\,.
\eeq
The Dirac action can be written in the various forms
\beqa
\frac{2i}{3}\int_M \bar{\psi} \star e\,e\,e\,D_{\omega}\psi +D_{\omega}\bar{\psi} \star e\,e\,e\,\psi &=& \frac{i}{12}\int_M  \epsilon_{IJKL}e^I \w e^J \w e^K \left(\bar{\psi} \g^L D_{\omega}\psi -D_{\omega}\bar{\psi} \g^L \psi \right) \nn\\
&=&\frac{i}{2}\int_M  \,e^\mu_I \left(\bar{\psi} \g^I \,D^\omega_{\mu} \psi -D^\omega_{\mu}\bar{\psi}\g^I \psi  \right) \, det(e) \,d^4x \,.
\eeqa
Similarly, the non-minimal Dirac coupling is
\beqa
\frac{2i}{3}\int_M\bar{\psi}\,e\,e\,e\,D_{\omega}\psi -D_{\omega}\bar{\psi}\,e\,e\,e\,\psi &=& -\frac{i}{12}\int_M \epsilon_{IJKL}\,e^I\,e^J\,e^K \left(\bar{\psi} \star \g^L D_{\omega} \psi + D_{\omega}\bar{\psi} \star \g^L \psi \right) \nn\\
&=& -\frac{i}{2}\int_M e^\mu_I \left(\bar{\psi} \star \g^I \,D^\omega_{\mu} \psi +D^\omega_{\mu}\bar{\psi}\star \g^I \psi  \right) \, det(e) \,d^4x\,.
\eeqa

\end{appendix}

\bibliography{MerRanBib}

\begin{thebibliography}{33}
\expandafter\ifx\csname natexlab\endcsname\relax\def\natexlab#1{#1}\fi
\expandafter\ifx\csname bibnamefont\endcsname\relax
  \def\bibnamefont#1{#1}\fi
\expandafter\ifx\csname bibfnamefont\endcsname\relax
  \def\bibfnamefont#1{#1}\fi
\expandafter\ifx\csname citenamefont\endcsname\relax
  \def\citenamefont#1{#1}\fi
\expandafter\ifx\csname url\endcsname\relax
  \def\url#1{\texttt{#1}}\fi
\expandafter\ifx\csname urlprefix\endcsname\relax\def\urlprefix{URL }\fi
\providecommand{\bibinfo}[2]{#2}
\providecommand{\eprint}[2][]{\url{#2}}

\bibitem[{\citenamefont{Ashtekar and Lewandowski}(2004)}]{AshLew04}
\bibinfo{author}{\bibfnamefont{A.}~\bibnamefont{Ashtekar}} \bibnamefont{and}
  \bibinfo{author}{\bibfnamefont{J.}~\bibnamefont{Lewandowski}}
  (\bibinfo{year}{2004}), \eprint{arXiv:gr-qc/0404018}.

\bibitem[{\citenamefont{Rovelli}(2004)}]{Rovelli:Book}
\bibinfo{author}{\bibfnamefont{C.}~\bibnamefont{Rovelli}},
  \emph{\bibinfo{title}{Quantum Gravity}} (\bibinfo{publisher}{Cambridge
  University Press}, \bibinfo{year}{2004}).

\bibitem[{\citenamefont{Thiemann}(2007)}]{Thi07}
\bibinfo{author}{\bibfnamefont{T.}~\bibnamefont{Thiemann}},
  \emph{\bibinfo{title}{Modern Canonical Quantum General Relativity}},
  Cambridge Monographs on Mathematical Physics (\bibinfo{publisher}{Cambridge
  University Press}, \bibinfo{year}{2007}), \bibinfo{edition}{1st} ed.

\bibitem[{\citenamefont{Holst}(1996)}]{Holst}
\bibinfo{author}{\bibfnamefont{S.}~\bibnamefont{Holst}},
  \bibinfo{journal}{Phys.Rev. D} \textbf{\bibinfo{volume}{53}},
  \bibinfo{pages}{5966} (\bibinfo{year}{1996}), \eprint{arXiv:gr-qc/9511026}.

\bibitem[{\citenamefont{Rovelli and Thiemann}(1998)}]{RovThi98}
\bibinfo{author}{\bibfnamefont{C.}~\bibnamefont{Rovelli}} \bibnamefont{and}
  \bibinfo{author}{\bibfnamefont{T.}~\bibnamefont{Thiemann}},
  \bibinfo{journal}{Phys. Rev.} \textbf{\bibinfo{volume}{D57}},
  \bibinfo{pages}{1009} (\bibinfo{year}{1998}), \eprint{gr-qc/9705059}.

\bibitem[{\citenamefont{Ashtekar et~al.}(2000)\citenamefont{Ashtekar, Baez, and
  Krasnov}}]{Ashtekar:entropy}
\bibinfo{author}{\bibfnamefont{A.}~\bibnamefont{Ashtekar}},
  \bibinfo{author}{\bibfnamefont{J.~C.} \bibnamefont{Baez}}, \bibnamefont{and}
  \bibinfo{author}{\bibfnamefont{K.}~\bibnamefont{Krasnov}},
  \bibinfo{journal}{Adv. Theor. Math. Phys.} \textbf{\bibinfo{volume}{4}},
  \bibinfo{pages}{1} (\bibinfo{year}{2000}), \eprint{arXiv:gr-qc/0005126}.

\bibitem[{\citenamefont{Rezende and Perez}(2009)}]{Perez:HolstAction}
\bibinfo{author}{\bibfnamefont{D.~J.} \bibnamefont{Rezende}} \bibnamefont{and}
  \bibinfo{author}{\bibfnamefont{A.}~\bibnamefont{Perez}},
  \bibinfo{journal}{Phys. Rev.} \textbf{\bibinfo{volume}{D79}},
  \bibinfo{pages}{064026} (\bibinfo{year}{2009}), \eprint{0902.3416}.

\bibitem[{\citenamefont{Gambini et~al.}(1999)\citenamefont{Gambini, Obregon,
  and Pullin}}]{GamObrPul99}
\bibinfo{author}{\bibfnamefont{R.}~\bibnamefont{Gambini}},
  \bibinfo{author}{\bibfnamefont{O.}~\bibnamefont{Obregon}}, \bibnamefont{and}
  \bibinfo{author}{\bibfnamefont{J.}~\bibnamefont{Pullin}},
  \bibinfo{journal}{Phys. Rev.} \textbf{\bibinfo{volume}{D59}},
  \bibinfo{pages}{047505} (\bibinfo{year}{1999}), \eprint{gr-qc/9801055}.

\bibitem[{\citenamefont{Mercuri}(2009)}]{Mer09p}
\bibinfo{author}{\bibfnamefont{S.}~\bibnamefont{Mercuri}}
  (\bibinfo{year}{2009}), \eprint{0903.2270}.

\bibitem[{\citenamefont{Date et~al.}(2009)\citenamefont{Date, Kaul, and
  Sengupta}}]{DatKauSen09}
\bibinfo{author}{\bibfnamefont{G.}~\bibnamefont{Date}},
  \bibinfo{author}{\bibfnamefont{R.~K.} \bibnamefont{Kaul}}, \bibnamefont{and}
  \bibinfo{author}{\bibfnamefont{S.}~\bibnamefont{Sengupta}},
  \bibinfo{journal}{Phys. Rev.} \textbf{\bibinfo{volume}{D79}},
  \bibinfo{pages}{044008} (\bibinfo{year}{2009}), \eprint{0811.4496}.

\bibitem[{\citenamefont{Mercuri}(2006{\natexlab{a}})}]{Mer06}
\bibinfo{author}{\bibfnamefont{S.}~\bibnamefont{Mercuri}},
  \bibinfo{journal}{Phys. Rev.} \textbf{\bibinfo{volume}{D73}},
  \bibinfo{pages}{084016} (\bibinfo{year}{2006}{\natexlab{a}}),
  \eprint{gr-qc/0601013}.

\bibitem[{\citenamefont{Mercuri}(2008)}]{Mer08}
\bibinfo{author}{\bibfnamefont{S.}~\bibnamefont{Mercuri}},
  \bibinfo{journal}{Phys. Rev.} \textbf{\bibinfo{volume}{D77}},
  \bibinfo{pages}{024036} (\bibinfo{year}{2008}), \eprint{0708.0037}.

\bibitem[{\citenamefont{Macdowell and Mansouri}(1977)}]{MMoriginal}
\bibinfo{author}{\bibfnamefont{S.}~\bibnamefont{Macdowell}} \bibnamefont{and}
  \bibinfo{author}{\bibfnamefont{F.}~\bibnamefont{Mansouri}},
  \bibinfo{journal}{Physical Review Letters} \textbf{\bibinfo{volume}{38}},
  \bibinfo{pages}{739} (\bibinfo{year}{1977}).

\bibitem[{\citenamefont{Stelle and West}(1979)}]{Stelle:1979va}
\bibinfo{author}{\bibfnamefont{K.~S.} \bibnamefont{Stelle}} \bibnamefont{and}
  \bibinfo{author}{\bibfnamefont{P.~C.} \bibnamefont{West}},
  \bibinfo{journal}{J. Phys.} \textbf{\bibinfo{volume}{A12}},
  \bibinfo{pages}{L205} (\bibinfo{year}{1979}).

\bibitem[{\citenamefont{Stelle and West}(1980)}]{Stelle:1979aj}
\bibinfo{author}{\bibfnamefont{K.~S.} \bibnamefont{Stelle}} \bibnamefont{and}
  \bibinfo{author}{\bibfnamefont{P.~C.} \bibnamefont{West}},
  \bibinfo{journal}{Phys. Rev.} \textbf{\bibinfo{volume}{D21}},
  \bibinfo{pages}{1466} (\bibinfo{year}{1980}).

\bibitem[{\citenamefont{Fukuyama}(1984)}]{Fukuyama:1984}
\bibinfo{author}{\bibfnamefont{T.}~\bibnamefont{Fukuyama}},
  \bibinfo{journal}{Annals of Physics} \textbf{\bibinfo{volume}{157}},
  \bibinfo{pages}{321} (\bibinfo{year}{1984}).

\bibitem[{\citenamefont{Randono}(2010{\natexlab{a}})}]{RandonodSSpaces}
\bibinfo{author}{\bibfnamefont{A.}~\bibnamefont{Randono}},
  \bibinfo{journal}{Class. Quant. Grav.} \textbf{\bibinfo{volume}{27}},
  \bibinfo{pages}{105008} (\bibinfo{year}{2010}{\natexlab{a}}),
  \eprint{0909.5435}.

\bibitem[{\citenamefont{Chamseddine and Mukhanov}(2010)}]{Chamseddine:2010rv}
\bibinfo{author}{\bibfnamefont{A.~H.} \bibnamefont{Chamseddine}}
  \bibnamefont{and} \bibinfo{author}{\bibfnamefont{V.}~\bibnamefont{Mukhanov}},
  \bibinfo{journal}{JHEP} \textbf{\bibinfo{volume}{03}}, \bibinfo{pages}{033}
  (\bibinfo{year}{2010}), \eprint{1002.0541}.

\bibitem[{\citenamefont{Randono}(2010{\natexlab{b}})}]{Randono:Condensate}
\bibinfo{author}{\bibfnamefont{A.}~\bibnamefont{Randono}}
  (\bibinfo{year}{2010}{\natexlab{b}}), \bibinfo{note}{in Preparation}.

\bibitem[{\citenamefont{Ashtekar}(1986)}]{Ash86-87a}
\bibinfo{author}{\bibfnamefont{A.}~\bibnamefont{Ashtekar}},
  \bibinfo{journal}{Phys. Rev. Lett.} \textbf{\bibinfo{volume}{57}},
  \bibinfo{pages}{2244} (\bibinfo{year}{1986}).

\bibitem[{\citenamefont{Ashtekar}(1987)}]{Ash86-87b}
\bibinfo{author}{\bibfnamefont{A.}~\bibnamefont{Ashtekar}},
  \bibinfo{journal}{Phys. Rev.} \textbf{\bibinfo{volume}{D36}},
  \bibinfo{pages}{1587} (\bibinfo{year}{1987}).

\bibitem[{\citenamefont{Barbero~G.}(1995{\natexlab{a}})}]{Bar95a}
\bibinfo{author}{\bibfnamefont{J.~F.} \bibnamefont{Barbero~G.}},
  \bibinfo{journal}{Phys. Rev.} \textbf{\bibinfo{volume}{D51}},
  \bibinfo{pages}{5498} (\bibinfo{year}{1995}{\natexlab{a}}),
  \eprint{gr-qc/9410013}.

\bibitem[{\citenamefont{Barbero~G.}(1995{\natexlab{b}})}]{Bar95b}
\bibinfo{author}{\bibfnamefont{J.~F.} \bibnamefont{Barbero~G.}},
  \bibinfo{journal}{Phys. Rev.} \textbf{\bibinfo{volume}{D51}},
  \bibinfo{pages}{5507} (\bibinfo{year}{1995}{\natexlab{b}}),
  \eprint{gr-qc/9410014}.

\bibitem[{\citenamefont{Eguchi et~al.}(1980)\citenamefont{Eguchi, Gilkey, and
  Hanson}}]{Eguchi:1980jx}
\bibinfo{author}{\bibfnamefont{T.}~\bibnamefont{Eguchi}},
  \bibinfo{author}{\bibfnamefont{P.~B.} \bibnamefont{Gilkey}},
  \bibnamefont{and} \bibinfo{author}{\bibfnamefont{A.~J.}
  \bibnamefont{Hanson}}, \bibinfo{journal}{Phys. Rept.}
  \textbf{\bibinfo{volume}{66}}, \bibinfo{pages}{213} (\bibinfo{year}{1980}).

\bibitem[{\citenamefont{Randono}(2005)}]{Ran05}
\bibinfo{author}{\bibfnamefont{A.}~\bibnamefont{Randono}}
  (\bibinfo{year}{2005}), \eprint{arXiv:hep-th/0510001}.

\bibitem[{\citenamefont{Mercuri}(2006{\natexlab{b}})}]{Mer06p}
\bibinfo{author}{\bibfnamefont{S.}~\bibnamefont{Mercuri}}
  (\bibinfo{year}{2006}{\natexlab{b}}), \eprint{gr-qc/0610026}.

\bibitem[{\citenamefont{Geroch}(1970)}]{ger70}
\bibinfo{author}{\bibfnamefont{R.~P.} \bibnamefont{Geroch}},
  \bibinfo{journal}{J. Math. Phys.} \textbf{\bibinfo{volume}{11}},
  \bibinfo{pages}{437} (\bibinfo{year}{1970}).

\bibitem[{\citenamefont{Ashtekar et~al.}(1989)\citenamefont{Ashtekar, Romano,
  and Tate}}]{AshRomTat89}
\bibinfo{author}{\bibfnamefont{A.}~\bibnamefont{Ashtekar}},
  \bibinfo{author}{\bibfnamefont{J.~D.} \bibnamefont{Romano}},
  \bibnamefont{and} \bibinfo{author}{\bibfnamefont{R.~S.} \bibnamefont{Tate}},
  \bibinfo{journal}{Phys. Rev.} \textbf{\bibinfo{volume}{D40}},
  \bibinfo{pages}{2572} (\bibinfo{year}{1989}).

\bibitem[{\citenamefont{Alexandrov}(2008)}]{Ale08}
\bibinfo{author}{\bibfnamefont{S.}~\bibnamefont{Alexandrov}},
  \bibinfo{journal}{Class. Quant. Grav.} \textbf{\bibinfo{volume}{25}},
  \bibinfo{pages}{145012} (\bibinfo{year}{2008}), \eprint{0802.1221}.

\bibitem[{\citenamefont{Chandia and Zanelli}(1997)}]{Zanelli:NiehYan}
\bibinfo{author}{\bibfnamefont{O.}~\bibnamefont{Chandia}} \bibnamefont{and}
  \bibinfo{author}{\bibfnamefont{J.}~\bibnamefont{Zanelli}},
  \bibinfo{journal}{Phys. Rev. D} \textbf{\bibinfo{volume}{55}},
  \bibinfo{pages}{7580} (\bibinfo{year}{1997}), \eprint{arXiv:hep-th/9702025}.

\bibitem[{\citenamefont{Randono and Wise}(2010)}]{RandonoWise}
\bibinfo{author}{\bibfnamefont{A.}~\bibnamefont{Randono}} \bibnamefont{and}
  \bibinfo{author}{\bibfnamefont{D.}~\bibnamefont{Wise}}
  (\bibinfo{year}{2010}), \bibinfo{note}{in Preparation}.

\bibitem[{\citenamefont{Wise}(2006)}]{Wise:MMgravity}
\bibinfo{author}{\bibfnamefont{D.~K.} \bibnamefont{Wise}}
  (\bibinfo{year}{2006}), \eprint{gr-qc/0611154}.

\bibitem[{\citenamefont{Wise}(2009)}]{Wise:2009fu}
\bibinfo{author}{\bibfnamefont{D.~K.} \bibnamefont{Wise}}
  (\bibinfo{year}{2009}), \eprint{0904.1738}.

\end{thebibliography}

\end{document}